%Paper: hep-th/9311014
%From: berglund@guinness.ias.edu (Per Berglund)
%Date: Tue, 2 Nov 93 14:47:36 EST
%Date (revised): Tue, 4 Jan 94 18:01:26 EST

%%%%%%%%%%%%%%%%%%%%%%%%%%%%%%%%%%%%%%%%%%%%%%%%%%%%%%%%%%%%%%%%%%%%%%
%%                                                                  %%
%%                             KDMS.TEX                             %%
%%                               by                                 %%
%%           Per Berglund                 Sheldon Katz              %%
%%                                and                               %%
%%       berglund@guinness.ias.edu      katz@math.okstate.edu	    %%
%%                                                                  %%
%%%%%%%%%%%%%%%%%%%%%%%%%%%%%%%%%%%%%%%%%%%%%%%%%%%%%%%%%%%%%%%%%%%%%%
\input harvmac%\input zip
%\draftmode
%%%%%%%%%%%%%%%%%%%%%%%%%%%%%%%%%%%%%%%%%%%%%%%%%%%%%%%%%%%%%%%%%%%%%%%%
%                                                                      %
%       "zip.tex", a set of macros to be used with "harvmac.tex"       %
%             Latest change: 18. X '92.   (Tristan Hubsch)             %
%                                                                      %
%%%%%%%%%%%%%%%%%%%%%%%%%%%%%%%%%%%%%%%%%%%%%%%%%%%%%%%%%%%%%%%%%%%%%%%%
 %
\catcode`@=11
\def\rlx{\relax\leavevmode}                  % Guess what this is for...
 %
 %
%%%%%%%%%%%%%%%%%%%%%%%%%%%%%%%%%%%%%%%%%%%%%%%%%%%%%%%%%%%%%%%%%%%%%%%%
%%%****FONTS****FONTS****FONTS****FONTS****FONTS****FONTS****FONTS****%%
 % That is, where fonts may not be available...
 %
 % Bold-face fonts in math
\font\tenmib=cmmib10
\font\sevenmib=cmmib10 at 7pt % =cmmib7 % if you have it
\font\fivemib=cmmib10 at 5pt  % =cmmib5 % if you have it
\font\tenbsy=cmbsy10
\font\sevenbsy=cmbsy10 at 7pt % =cmbsy7 % if you have it
\font\fivebsy=cmbsy10 at 5pt  % =cmbsy5 % if you have it
\def\BMfont{\textfont0\tenbf \scriptfont0\sevenbf
                              \scriptscriptfont0\fivebf
            \textfont1\tenmib \scriptfont1\sevenmib
                               \scriptscriptfont1\fivemib
            \textfont2\tenbsy \scriptfont2\sevenbsy
                               \scriptscriptfont2\fivebsy}
\def\BM#1{\rlx\ifmmode\mathchoice
                      {\hbox{$\BMfont#1$}}
                      {\hbox{$\BMfont#1$}}
                      {\hbox{$\scriptstyle\BMfont#1$}}
                      {\hbox{$\scriptscriptstyle\BMfont#1$}}
                 \else{$\BMfont#1$}\fi}
 %
 % If you don't have the above fonts, comment out the above 21 lines,
 % >> send two pounds of live cockroaches to your TeX distributor <<
 % and use the Poor man's boldface (a la D. Knuth, and a bit better):
 %
 %\def\BM#1{\relax\leavevmode\setbox0=\hbox{$#1$}
 %           \kern-.025em\copy0\kern-\wd0
 %            \kern.05em\copy0\kern-\wd0
 %             \kern-.025em\raise.0433em\copy0\kern-\wd0
 %              \raise.0144em\box0 }
 %
 % Some basic black-board bold (capital) letters
 % ...should work rather well in sub- and super-scripts also...
 %
\def\inbar{\vrule height1.5ex width.4pt depth0pt}
\def\sinbar{\vrule height1ex width.35pt depth0pt}
\def\ssinbar{\vrule height.7ex width.3pt depth0pt}
\font\cmss=cmss10
\font\cmsss=cmss10 at 7pt
\def\ZZ{\rlx\leavevmode
             \ifmmode\mathchoice
                    {\hbox{\cmss Z\kern-.4em Z}}
                    {\hbox{\cmss Z\kern-.4em Z}}
                    {\lower.9pt\hbox{\cmsss Z\kern-.36em Z}}
                    {\lower1.2pt\hbox{\cmsss Z\kern-.36em Z}}
               \else{\cmss Z\kern-.4em Z}\fi}
\def\Ik{\rlx{\rm I\kern-.18em k}}  % Yes, I know. This ain't capital.
\def\IC{\rlx\leavevmode
             \ifmmode\mathchoice
                    {\hbox{\kern.33em\inbar\kern-.3em{\rm C}}}
                    {\hbox{\kern.33em\inbar\kern-.3em{\rm C}}}
                    {\hbox{\kern.28em\sinbar\kern-.25em{\sevenrm C}}}
                    {\hbox{\kern.25em\ssinbar\kern-.22em{\fiverm C}}}
             \else{\hbox{\kern.3em\inbar\kern-.3em{\rm C}}}\fi}
\def\IP{\rlx{\rm I\kern-.18em P}}
\def\IR{\rlx{\rm I\kern-.18em R}}
\def\Ione{\rlx{\rm 1\kern-2.7pt l}}
 %
%%%%%%%%%%%%%%%%%%%%%%%%%%%%%%%%%%%%%%%%%%%%%%%%%%%%%%%%%%%%%%%%%%%%%%%%
%%%****SHAPE****SHAPE****SHAPE****SHAPE****SHAPE****SHAPE****SHAPE****%%
 %
 % Get in shape, Man, (never mind the content)!

 %

\def\intem#1{\par\leavevmode%
              \llap{\hbox to\parindent{\hss{#1}\hfill~}}\ignorespaces}
 %
 % Indents #1 lines by width of #2 and puts #2 in the "niche".

 % Similar to "niche" :

 %
 % Math...
 % My version of \eqalign, \eqalignno ...
\newskip\humongous \humongous=0pt plus 1000pt minus 1000pt   % isn't it?
\def\caja{\mathsurround=0pt}
\newif\ifdtup
 %
 % display pattern: [         a &= b        \cr]
\def\eqalign#1{\,\vcenter{\openup2\jot \caja
     \ialign{\strut \hfil$\displaystyle{##}$&$
      \displaystyle{{}##}$\hfil\crcr#1\crcr}}\,}
 %
 % display pattern: [   a &= b  &  c &= d   \cr]

 %
 % display to full hsize, numbered at far right

 %
 % display pattern: [         a &= b        &(*) \cr]

 %
 % display pattern: [      a &= b &= c      &(*) \cr]

 %
 % display pattern: [   a &= b  &  c &= d   &(*) \cr]

 %
 % display pattern: [    a &= b &= c &= d   &(*) \cr]

 %
 % For extra v-space between rows of a matrix or eqn-alignment,
 % use "\noalign{\vskip2mm}". In the above equation alignments,
 % "\openup2mm" does the same, but has no effect in "\matrix".
 %
%%%%%%%%%%%%%%%%%%%%%%%%%%%%%%%%%%%%%%%%%%%%%%%%%%%%%%%%%%%%%%%%%%%%%%%%
%%%****SHORT****SHORT****SHORT****SHORT****SHORT****SHORT****SHORT****%%
 %
 % Redefinitions of TeX's commands :
 %
          % Polish l-slash, L-slash
        % Scandinavian o-slash, O-slash
          % P-mirror, double-S
        % tie-after, cedilla
\let\ii=\i          % include mathmode !!!
          % under-bar, under-dot
\def\,{\hskip1.5pt}           % why only in math-mode?
 %
 % Some abbreviations that save typing :
\let\a=\alpha

\let\c=\chi
\let\d=\delta                    \let\D=\Delta
\let\e=\epsilon     
\let\f=\phi                       
\let\g=\gamma                                    \let\G=\Gamma

\let\i=\iota
\let\j=\psi                                      

\let\l=\lambda

\let\p=\pi          \let\vp=\varpi               
                   \let\Q=\Theta
\let\r=\rho         
\let\s=\sigma                   

                                    \let\W=\Omega

 %
 % Additional math symbols
 %
\def\Box{\sqcap\llap{$\sqcup$}}
\def\lapp{\lower.4ex\hbox{\rlap{$\sim$}} \raise.4ex\hbox{$<$}}
\def\gapp{\lower.4ex\hbox{\rlap{$\sim$}} \raise.4ex\hbox{$>$}}
\def\con{\ifmmode\raise.1ex\hbox{\bf*}
          \else\raise.1ex\hbox{\bf*}\fi}
\def\bo{{\raise.15ex\hbox{\large$\Box\kern-.39em$}}}
  %  a very fat nothing

\def\dual{\relax\leavevmode\lower.9ex\hbox{\titlerms*}}
\def\define{\buildrel\rm def\over =}

\let\8=\otimes
 %
 %
%%%%%%%%%%%%%%%%%%%%%%%%%%%%%%%%%%%%%%%%%%%%%%%%%%%%%%%%%%%%%%%%%%%%%%%%
%%****MACROS***MACROS***MACROS***MACROS***MACROS***MACROS***MACROS****%%
 %
 % Math macros
 %

\let\2=\underline

 %
 % Let's take arguments...
%  Use:  "A\like{B}".
\def\dt#1{{\buildrel{\smash{\lower1pt\hbox{.}}}\over{#1}}}

\font\eightrm=cmr8
\def\6(#1){\relax\leavevmode\hbox{\eightrm(}#1\hbox{\eightrm)}}
\def\0#1{\relax\ifmmode\mathaccent"7017{#1}     % a little circle atop,
                \else\accent23#1\relax\fi}      % as a halo of a saint
\def\7#1#2{{\mathop{\null#2}\limits^{#1}}}      % puts #1 atop #2
\def\5#1#2{{\mathop{\null#2}\limits_{#1}}}      % puts #1 beneath #2
 %
 % Will grow vertically with size of argument

 %
 % Vertical arrows with labels

 %
 % For vert. arrows to grow, say "\bigg\down\crlap{...}", using the
 % side-script: a label for vertical delimiters

 %

 %
 % Horizontal arrows that can grow
\newbox\t@b@x
\def\rightarrowfill{$\m@th \mathord- \mkern-6mu
     \cleaders\hbox{$\mkern-2mu \mathord- \mkern-2mu$}\hfill
      \mkern-6mu \mathord\rightarrow$}
\def\tooo#1{\setbox\t@b@x=\hbox{$\scriptstyle#1$}%
             \mathrel{\mathop{\hbox to\wd\t@b@x{\rightarrowfill}}%
              \limits^{#1}}\,}
\def\leftarrowfill{$\m@th \mathord\leftarrow \mkern-6mu
     \cleaders\hbox{$\mkern-2mu \mathord- \mkern-2mu$}\hfill
      \mkern-6mu \mathord-$}
\def\froo#1{\setbox\t@b@x=\hbox{$\scriptstyle#1$}%
             \mathrel{\mathop{\hbox to\wd\t@b@x{\leftarrowfill}}%
              \limits^{#1}}\,}
 %
 % fractions
\def\frac#1#2{{#1\over#2}}
\def\frc#1#2{\relax\ifmmode{\textstyle{#1\over#2}} % A small fraction,
                    \else$#1\over#2$\fi}           % good in text.
                            % Like {1\over{#1}}
 %
 % The basic Theorem-like macro, uses equation numbers
 % Use:  \Claim\cLABEL{Theorem}{This is a theorem.}
\def\Claim#1#2#3{\bigskip\begingroup%
                  \xdef #1{\secsym\the\meqno}%
                   \writedef{#1\leftbracket#1}%
                    \global\advance\meqno by1\wrlabeL#1%
                     \noindent{\bf#2}\,#1{}\,:~\sl#3\vskip1mm\endgroup}

\def\QED{\rlx\hfill$\Box$\kern-7pt\raise3pt\hbox{$\surd$}\bigskip}
 %
 % Math miscellanea
 %

 %
\def\muthstrut{\vphantom1}
\def\mutrix#1{\null\,\vcenter{\normalbaselines\m@th
        \ialign{\hfil$##$\hfil&&~\hfil$##$\hfill\crcr
            \muthstrut\crcr\noalign{\kern-\baselineskip}
            #1\crcr\muthstrut\crcr\noalign{\kern-\baselineskip}}}\,}

 %
 % Young tableaux: use "\Box" for a box and "\Z" for newline
 % The 2-1 stair-tableau:  \YT{\Box\Box}{\Box\Box\Z\Box}
 % Note: the first argument sets the width!
\def\YT#1#2{\vcenter{\hbox{\vbox{\baselineskip0pt\parskip=\medskipamount%
             \def\Box{$\sqcap\llap{$\sqcup$}$\kern-1.2pt}%
              \def\Z{\hfil\vskip-5.8pt}\lineskiplimit0pt\lineskip0pt%
               \setbox0=\hbox{#1}\hsize\wd0\parindent=0pt#2}\,}}}
\def\EU{\rlx\ifmmode \c_{{}_E} \else$\c_{{}_E}$\fi}
\def\TM{\rlx\ifmmode {\cal T_M} \else$\cal T_M$\fi}
\def\TW{\rlx\ifmmode {\cal T_W} \else$\cal T_W$\fi}
\def\CM{\rlx\ifmmode {\cal T\rlap{\bf*}\!\!_M}
             \else$\cal T\rlap{\bf*}\!\!_M$\fi}
\def\hm#1#2{\rlx\ifmmode H^{#1}({\cal M},{#2})
                 \else$H^{#1}({\cal M},{#2})$\fi}
\def\CP#1{\rlx\ifmmode\IP^{#1}\else\IP$^{#1}$\fi}
\def\cP#1{\rlx\ifmmode\IC{\rm P}^{#1}\else$\IC{\rm P}^{#1}$\fi}
\def\WCP#1#2{\rlx\IP^{#1}_{#2}}
\def\sll#1{\rlx\rlap{\,\raise1pt\hbox{/}}{#1}}
\def\Sll#1{\rlx\rlap{\,\kern.6pt\raise1pt\hbox{/}}{#1}\kern-.6pt}
%

 %
 % Text miscellanea
 %
\def\ie{\hbox{\it i.e.}}        % By Knuth, use commas: ..., \ie, ... !

\def\CY{Calabi-\kern-.2em Yau}

\def\3{\ifmmode\ldots\else$\ldots$\fi}
\def\Z{\hfil\break\rlx\hbox{}\quad}
\def\3{\ifmmode\ldots\else$\ldots$\fi}
\def\?{d\kern-.3em\raise.64ex\hbox{-}}           % d-dash
\def\9{\raise.43ex\hbox{-}\kern-.37em D}         % D-Dash

 %
 % References
 %
\def\I#1{{\it ibid.\,}{\bf#1\,}}

\def\pre#1{{\it University of #1 report}}

\def\NP#1{{\it Nucl.\,Phys.\,}{\bf#1\,}}
\def\PL#1{{\it Phys.\,Lett.\,}{\bf#1\,}}

\def\MPL#1{{\it Mod.\,Phys.\,Lett.\,}{\bf#1\,}}

\def\CMP#1{{\it Commun.\,Math.\,Phys.\,}{\bf#1\,}}

 %
 %
%%%%%%%%%%%%%%%%%%%%%%%%%%%%%%%%%%%%%%%%%%%%%%%%%%%%%%%%%%%%%%%%%%%%%%%%
%%============>>>>            SAVE  TIMBER            <<<<============%%
 %
\baselineskip=13.0861pt plus2pt minus1pt
\parskip=\medskipamount
\let\ft=\foot
\noblackbox
\def\SaveTimber{\abovedisplayskip=1.5ex plus.3ex minus.5ex
                \belowdisplayskip=1.5ex plus.3ex minus.5ex
                \abovedisplayshortskip=.2ex plus.2ex minus.4ex
                \belowdisplayshortskip=1.5ex plus.2ex minus.4ex
                \baselineskip=12pt plus1pt minus.5pt
 \parskip=\smallskipamount
 \def\ft##1{\unskip\,\begingroup\footskip9pt plus1pt minus1pt\setbox%
             \strutbox=\hbox{\vrule height6pt depth4.5pt width0pt}%
              \global\advance\ftno by1\footnote{$^{\the\ftno)}$}{##1}%
               \endgroup}
 \def\listrefs{\footatend\vfill\immediate\closeout\rfile%
                \writestoppt\baselineskip=10pt%
                 \centerline{{\bf References}}%
                  \bigskip{\frenchspacing\parindent=20pt\escapechar=` %
                   \rightskip=0pt plus4em\spaceskip=.3333em%
                    \input refs.tmp\vfill\eject}\nonfrenchspacing}}
 %
 % For European standard
\def\Afour{\ifx\answ\bigans
            \hsize=16.5truecm\vsize=24.7truecm
             \else
              \hsize=24.7truecm\vsize=16.5truecm
               \fi}
\catcode`@=12
%%%%%%%%%%%%%%%%%%%%%%%%%%%%%%%%%%%%%%%%%%%%%%%%%%%%%%%%%%%%%%%%%%%%%%%%
%                                                                      %
%                          End of  "zip.tex".                          %
%                        May the article begin!                        %
%                                                                      %
%%%%%%%%%%%%%%%%%%%%%%%%%%%%%%%%%%%%%%%%%%%%%%%%%%%%%%%%%%%%%%%%%%%%%%%%

\baselineskip=12pt
%\SaveTimber
%\proofmodetrue
%\pageno=1 \chapternumberstrue
%\chapno=0 \figurechapternumberstrue \tablechapternumberstrue
%\forwardreferencetrue
%%\initialeqmacro
%\noblackbox

\def\LG{Landau-Ginzburg}
\def\WP{\rlx{W\rm I\kern-.18em P}}
\def\ph{\phantom-}
\def\cp#1#2{\hbox{$\IP_{#1}^{#2}$}}
\def\wp{\cp{(k_1,k_2,k_3,k_4,k_5)}{4}}
\def\cM{\cal M}
\def\Mhat{\hat{\cal M}}
\def\Mbar{\bar{\cal M}}

\def\Wbar{\bar{\cal W}}
\def\CY{Calabi-Yau}
\def\Hom{\rm Hom}
\def\Np{{\cal N}(P)}
\def\PN{P^{\circ}\cap N}

 \def\Xb{\relax\leavevmode\hbox{$X$\kern-.6em%
                   \vrule height.4pt width5.7pt depth-1.8ex}\kern1pt}
 \def\rd{{\rm d}}

     %

 % Save 16.67% of timber
%\baselineskip=13.6 pt plus 2pt minus1pt
\def\Afour{\hsize=16.5truecm\vsize=24.7truecm}
% \Afour
%\leftline{\hsize=36mm\font\WARN=cmr7\font\warN=cmr5
%               \boxit{2pt}{\vskip-3pt\baselineskip=7pt\noindent\WARN
%                P{\warN LEASE}, {\warN DO NOT CIRCULATE}:
%       This is merely a tentative first sketch.\vskip0pt}}\vskip-10.1mm
\Title{\vbox{\baselineskip12pt \hbox{IASSNS-HEP-93/65}
                                  \hbox{NSF-ITP-93-128}
                                  \hbox{OSU-M-93-2}
                                  \hbox{UTTG-24-93}}}
{\vbox{\centerline{Mirror Symmetry for Hypersurfaces in Weighted}\vskip1mm
            \centerline{Projective Space and Topological Couplings}}}
\centerline{\titlerms Per Berglund} \vskip 1mm
\centerline{\it School of Natural Science} \vskip0mm
\centerline{\it Institute for Advanced Study}       \vskip0mm
 \centerline{\it Olden Lane}  \vskip0mm
 \centerline{\it Princeton, NJ 08540}                 \vskip0mm
 \centerline{\rm berglund\,@\,guinness.ias.edu}              \vskip 1	mm
\vskip .2in
 \centerline{\titlerms and}
\vskip .2in
\centerline{\titlerms Sheldon Katz}    \vskip1mm
 \centerline{\it Department of Mathematics}                   \vskip0mm
 \centerline{\it Oklahoma State University} \vskip0mm
 \centerline{\it Stillwater, OK~74078} \vskip0mm
 \centerline{\rm katz\,@\,math.okstate.edu}
\vfill

\centerline{ABSTRACT}\vskip2mm
\vbox{\narrower\narrower\narrower\baselineskip=12pt\noindent
By means of toric geometry we study hypersurfaces in weighted projective
space of dimension four. In particular we compute for a given manifold its
intrinsic topological coupling. We find that the result agrees with the
calculation of the corresponding coupling on the mirror model in the
large complex structure limit.}

\Date{\vbox{ \line{December 1993 \hfill}}}
\footline{\hss\tenrm--\,\folio\,--\hss}

\nref\rgp{B.R.~Greene and R.~Plesser: ``An Introduction to Mirror Manifolds'',
in {\it Essays on Mirror
       Symmetry}, and references therein, ed.\ S.-T.~Yau  (Intl.\ Press,
Hong Kong, 1992).}

\nref\rLance{For a review and references, see L.~Dixon: in {\it
     Superstrings, Unified Theories and Cosmology 1987},
     eds.~G.~Furlan et al.\ (World Scientific, Singapore, 1988)
     \,p.~67--127.}

\nref\rMax{M.~Kreuzer and H.~Skarke: \CMP{150} (1992) 137.}

\nref\rAR{A.~Klemm and R.~Schimmrigk~: ``Landau-Ginzburg String Vacua'',
{\it CERN preprint} CERN-TH 6459/92\semi
          M.~Kreuzer and H.~Skarke: \NP{B388} (1992) 113.}

\nref\rMaxII{M.~Kreuzer and H.~Skarke:
``All Abelian Symmetries of Landau-Ginzburg Potentials'', CERN-TH-6705/92.}

\nref\rCdGP{P.~Candelas, X.~ de la Ossa, P.~Green and L.~Parkes: \NP{B359}~
(1991)~21.}

\nref\rPicF{D.R.~Morrison: ``Picard-Fuchs Equations and Mirror Maps For
Hypersurfaces'', in {\it Essays on Mirror
       Symmetry}, p.1, ed.\ S.-T.~Yau  (Intl.\ Press,
Hong Kong, 1992)\semi
          A.~Font: \NP{B391}(1993)358\semi
A.~Klemm and S.~Theisen: ``Mirror maps and instanton sums for complete
intersections in weighted projective space'', Munich University preprint
LMU-TPW-93-08, \NP{B389}(1993)153\semi
A.~Libgober and J.~Teitelbaum: `` Lines on \CY\ complete
       intersections, mirror symmetry and Picard-Fuchs equations''
       \pre{Illinois} (1992).}

\nref\rCdFKM{P.~Candelas, X.~de la Ossa, A.~Font, S.~Katz,
D.R.~Morrison: ``Mirror Symmetry for Two Parameter Models - I'',
 CERN-TH.\ 6884/93,
UTTG-15-93, NEIP-93-005, OSU-M-93-1.}

\nref\rHKTY{S.~Hosono, A.~Klemm, S.~Theisen and S.-T.~Yau: ``Mirror Symmetry,
Mirror Map and Applications to Calabi-Yau Hypersurfaces'', HUTMP-93/0801.}

\nref\rSL{T.~H\"ubsch and S.-T.~Yau:
       Mod.~Phys.~Lett.{\bf~A7}(1992)3277.}

\nref\rBatdual{V.~Batyrev: ``Dual polyhedra and mirror symmetry for Calabi-Yau
hypersurfaces in toric varieties'', Preprint, November 18, 1992.}

\nref\rGriff{P.A.~Griffiths: {\it Ann. of Math.} (2){\bf 90}(1969).}

\nref\rFerrara{For a review see S.~Ferrara: \MPL{A6} (1991) 2175 and
references therein.}

\nref\rDKL{L.~Dixon, V.~Kaplunovsky and J.~Louis: \NP{B329} (1990)
27.}

\nref\rBvS{V.~Batyrev and D.~van Straten: ``Generalized Hypergeometric
Functions and Rational Curves on \CY\ Complete Intersections in Toric
Varieties'',  Preprint.}

\nref\rAGM{P.S.~Aspinwall, B.R.~Greene and D.R.~Morrison:
       \PL{B303} (1993) 249,
       ``Calabi-Yau Moduli Space, Mirror Manifolds and Spacetime
         Topology Change in String Theory'',
       {\it Institute for Advanced Study report} IASSNS-HEP-93/38.}

\nref\rDSWW{M.~Dine, N.~Seiberg, X.G.~Wen and E.~Witten:
       \NP{B278}(1986)769, \I{B289}(1987)319.}

\nref\rGMC{P.~Berglund and T.~H\"ubsch: \NP{B393}(1993)377, also in the Mirror
Symmetry Workshop Proceedings, MSRI, Berkeley, May 1991.}

\nref\rBG{R.~Bryant and P.~Griffiths, {\it Progress in Mathematics} {\bf 36}
p. 77, (Birkh\"auser, Boston, 1983).}

\nref\rRoll{P.~Candelas, T.~H\"ubsch and P.~Green: \NP{B330} (1990) 49.}

\nref\rCO{P.~Candelas and X.~de la Ossa: \NP{B355}(1991)455.}

\nref\rGriffII{P.A.~Griffiths: {\it Bull.\,Amer.\,Math.\,Soc.\,}
{\bf 76} (1970) 228.}

\nref\rCF{A.C.~Cadavid and S.~Ferrara: \PL{B267}(1991)193.}

\nref\rLSW{W.~Lerche, D.~J.~Smit and N.~P.~Warner: \NP{B372} (1992) 87.}

\nref\rCDFLL{A.~Ceresole, R.~D'Auria, S.~Ferrara, W.~Lerche and J.~Louis:
{\it Int.J.Mod.Phys.} {\bf A8} (1993)79.}

\nref\rPeriod{P.~Berglund, P.~Candelas, X.~de~la~Ossa, A.~Font, T.~H\"ubsch,
D.~Jan\v ci\'c and F.~Quevedo,
``Periods for Calabi-Yau and Landau-Ginzburg Vacua'',
CERN-TH. 6865/93, HUPAPP-93/3,NEIP 93-004, NSF-ITP-93-96, UTTG-13-93.}

\nref\rmdmm{P.S.~Aspinwall, B.R.~Greene and D.R.~Morrison: ``The
Monomial-Divisor Mirror Map'', IASSNS-HEP-93/43.}

\nref\rFul{W.~Fulton: Introduction to toric varieties,
Annals of Math.\ Studies, vol.~131,
Princeton University Press, Princeton, 1993.}

\nref\rCOK{P.~Candelas, X.~de~la~Ossa and S.~Katz: ``Mirror Symmetry for
Calabi-Yau Hypersurfaces in Weighted $\IP^4$ and an Extension of
Landau-Ginzburg Theory'', in preparation.}

\nref\rBatyrev{V.V.~Batyrev: {\it Duke Math.\,Journ.\,} {\bf 69} (1993) 349.}

\nref\rBK{P.~Berglund and S.~Katz: in preparation.}

\nref\rOP{T.~Oda and H.S.~Park: {\it T\^ohoku Math.~J.}
(2) {\bf 43} (1991), 375.}

\nref\rLeVaWa{W.~Lerche, C.~Vafa and N.~Warner: \NP{B324} (1989) 427\semi
              P.~Candelas:\NP{B298} (1988) 458.}

\nref\rVTop{C.~Vafa~: Mod. Phys. Lett. {\bf A6} (1991) 337.}

\nref\rGH{P.~Green and T.~H\"ubsch: \CMP{113} (1987) 505.}

\nref\rBatquant{V.~Batyrev: ``Quantum Cohomology Rings of Toric Manifolds'',
Preprint.}

\nref\rdoltor{I.~Dolgachev:  ``Weighted projective varieties'', in {\it
Group Actions and Vector Fields}, ed. J.B. Carrell, pages
  34--71. Springer-Verlag, Berlin-Heidelberg-New York, 1982.
 Proceedings, Vancouver 1981.}

\newsec{Introduction}
\noindent
Mirror symmetry \rgp\ is in one respect the result of a very simple
observation; at the level of the $(2,2)$ superconformal field theory the
relative sign of the $U(1)$-current is ambiguous and the two different
choices lead to isomorphic theories \rLance. On the other hand, when the theory
is formulated as a two-dimensional $N=2$ supersymmetric non-linear $\s$-model
on
a \CY\ manifold as the target space the result is far from trivial. The
change of  sign of the $U(1)$-current interchanges the $(c,c)$ and the
$(a,c)$ ring and so the effect is  to flip the sign of the Euler number,
$\EU=2(b_{1,1}-b_{2,1})$,
thus relating two topologically distinct manifolds. This has far-reaching
consequences.
In particular by computing the $(2,1)$ form couplings on ${\cal W}$, the mirror
of ${\cal M}$, we obtain, by mirror symmetry, the fully corrected $(1,1)$ form
couplings on ${\cal M}$. This statement is true for all of moduli space and
hence we know the `quantum' K\"ahler couplings not only in the large radius
limit but at any point in the space of K\"ahler deformations of ${\cal M}$.

Based on the recent classification efforts of $N=2$ string vacua
\refs{\rMax,\rAR,\rMaxII}
the class of known $(2,2)$ Landau-Ginzburg orbifolds is
(almost) mirror symmetric. Still mirror symmetry is merely a conjecture
and the number of models for which mirror symmetry has been explicitly checked
is rather small \refs{\rCdGP,\rPicF,\rCdFKM,\rHKTY}~.
In this paper we continue the effort of verifying mirror symmetry. Following
Candelas {\it et al.} \rCdGP\
we study a one-dimensional subspace of the space of
complex deformations of a class of manifolds ${\cal W}$ whose mirrors are
transverse hypersurfaces $\cM$ in weighted $\cp{(k_1,\ldots,k_5)}{4}$. We
restrict to the deformation which corresponds
to the K\"ahler form inherited from
the ambient space of $\cM$.\foot{In general, due to the quotient
singularities from the projectivization, it is $k$ times the na\"\ii ve
K\"ahler form
which is well-defined. The weight $k$ of the hyperplane class of
${\cal M}$ is a natural
number  which depends on the
singularity, see section~2 and appendix~A.}
 It has been argued that such a deformation always exists
for a generic choice of the defining polynomial for ${\cal W}$ \rSL,
and we will show that this is indeed the case, using the construction of
mirror manifolds proposed by Batyrev \rBatdual.
With this information at hand we find the
periods  which are solutions of the Picard-Fuchs equation~\rGriff. The
existence
of this equation follows from the fact that the moduli space is not
just K\"ahler  but `special K\"ahler' due to the $N=2$ world-sheet
supersymmetry~\refs{\rFerrara,\rDKL}.  The knowledge of the periods allows us
to solve the theory completely. In particular, we calculate  the Yukawa
coupling as a function on our one parameter subspace of the moduli space. In
the large complex structure
limit this will give predictions for the topological K\"ahler  coupling
$y_{JJJ}$  on
${\cal M}$. Indeed, this is confirmed by an intersection calculation where
we find that the couplings are exactly the same
(see~(2.4)
 and~(5.14)). This latter
unexpected agreement suggests that it may be possible to formulate the
computation only in terms of the K\"ahler modulus dual to the fundamental
monomial.
In fact, once the fundamental period has been found we may think of
the computation as carried out on the K\"ahler moduli space of $\cM$
rather
than on the space of complex deformations of ${\cal W}$. Special geometry is
a property not just of the complex deformations but for the K\"ahler
deformations as well. Hence, one might expect that when formulating the theory
completely on the K\"ahler side the same Picard-Fuchs equation will
appear. This would not only allow us to solve the theory completely within
$\cM$ but it would also give a proof of mirror symmetry for all of moduli
space.  See also \rBvS.

The paper is organized as follows. In section~2 we
 derive an expression for the topological Yukawa coupling $y_{JJJ}$ on
$\cM$. We then discuss some basic properties of toric geometry needed
to compute the period (section~3) as well as to understand the complex
structure moduli space
of the ${\cal W}$ (section~4).
In section~5  the Yukawa coupling $y_{\j\j\j}^{\scriptstyle {\rm lcs}}$
in the large
complex structure limit on ${\cal W}$ is obtained from the
knowledge of the monodromy
around $\j=\infty$.
Comparison with $y_{JJJ}$ on $\cM$ gives complete agreement for all
models studied so far. Our conclusions and
discussions are left for section~6 while some technical details
regarding the analytic continuation of the periods and the derivation
of the weight $k$ of the hyperplane class of
${\cal M}$ can be found in  the appendices.

\newsec{The Topological Yukawa Coupling}
\noindent
Let $p(x_i)=0$ define a  hypersurface
 ${\cal M} \in \cp{(k_1,k_2,k_3,k_4,k_5)}{4}[d]$
 where $p(x_i)$ is a quasi homogeneous
polynomial of degree
$d=\sum_ik_i$.\foot{As usual $\cp{(k_1,k_2,k_3,k_4,k_5)}{4}[d]$
refers to the family
of degree $d$ hypersurfaces in the weighted projective space
$\cp{(k_1,k_2,k_3,k_4,k_5)}{4}$.}
Recall that the $x_i$ are weighted coordinates,
\eqn\eweight{
x_i\sim \l^{k_i} x_i\,~i=1,\ldots,5\,;
\qquad p(\l^{k_i} x_i)\sim \l^d p(x_i)\,.
}
Associated to $\cM$ is the parameter space of K\"ahler deformations and
complex structure deformations of dimension
$h_{1,1}=\dim H^{1,1}({\cal M})$ and
$h_{2,1}=\dim H^{2,1}({\cal M})$ respectively\foot{We actually mean the
dimensions of the spaces of harmonic $(1,1)$ and $(2,1)$ forms on an
appropriate Calabi-Yau desingularization
of $\cM$.}.
We are interested in computing the
$(1,1)$ form Yukawa coupling restricted to the case when
all three $(1,1)$ forms
are equal to the K\"ahler form $J\in H^{1,1}({\cal M})$
inherited from the ambient space,
\eqn\eYUK{y_{JJJ}~=~\int_{{\cal M}} J^3\quad =\quad H^3,}
where $H\in H^2({\cal M})$ is the hyperplane class and $H^3$ is the triple
intersection number of $H$.
This coupling is topological, \ie\ it does not depend on the choice of complex
structure of ${\cal M}$; neither do any of the other $(1,1)$-couplings.
In the light of the recent results on topology change in string
theory~\rAGM\ one may worry that the intersection number will depend on
which large radius limit is chosen.
However, the topology change arises from choosing different desingularizations
of $\cM$, and since $H^3$ can be calculated on any of these by pulling back
the calculation for $\cM$, the intersection number is well defined.
In contrast to the $(2,1)$ form coupling, $y_{JJJ}$ will receive quantum
corrections in the form of instanton contributions \rDSWW.
However, for large radius this contribution is exponentially suppressed
and $y_{JJJ}$ will give the correct result of the corresponding three-point
function in the underlying conformal field theory. Thus, in this limit, we have
the possibility of comparing calculations made on ${\cal M}$ and ${\cal W}$.

For those models in which the projectivity constraint \eweight\
does not lead to any
singularities on the manifold, hence no extra $(1,1)$ forms,
\eqn\eyukno{y_{JJJ}~=~ {d\over \prod_{i=1}^5 k_i}~.}
However, eq.~\eyukno\ holds only for four hypersurfaces in weighted
$\CP{4}$. Rather, there are singularities on the manifold which
when blown up contribute to the number of $(1,1)$ forms and also change
\eyukno. As is shown in appendix~A, the effect on $y_{JJJ}$
leads to a very simple correction of \eyukno,
\eqn\eyuksing{y_{JJJ}~=~ {d\, k^3\over \prod_{i=1}^5 k_i}~,}
where $k$, the weight of the hyperplane class of
${\cal M}$, is a natural number.
In appendix~A we give a derivation of $k$
for any hypersurface in a weighted projective space of dimension four.
Below we will, following~\rCdFKM, compute $y_{JJJ}$ explicitly for two
examples as well as use \eyuksing\ for comparison.  Related intersection
numbers have been computed in \rHKTY.

\noindent
{\bf Example 1.} Consider the hypersurface
${\cal M}_1\in\cp{(1,1,1,2,2)}{4}[7]$ defined
by $p_1=0$ where $p_1$ is a transverse polynomial. Note that $p_1$ cannot be
a Fermat.  There is a $\ZZ_2$ fixed point set given by a $\IP^1$. From~(A.3)
we have $k=2$ for this example.  Let
$H$ be the linear system of quadratic polynomials.
We are interested in computing the triple intersection
$H^3$. Since $x_4$ and $x_5$ have degree $2$, both $x_4={\rm const}$ and
$x_5={\rm const}$ are in
$H$. Thus, in the patch $x_1=1$ we can describe $H^3$   by the intersection of
$x_4={\rm const}$, $x_5={\rm const}$ and a quadratic equation in $x_2$ and
$x_3$.
If we choose $p_1=x_1^7 +x_2^7 + x_2 x_4^3 + x_3^7 + x_3 x_5^3$ we get
$x_2^7 + x_3^7 = {\rm const}$
which together with the quadratic equation in $x_2$ and $x_3$ form a system
of equations with
$2\cdot  7=14$ solutions. Thus,
\eqn\eXXX{
          H^3~=~14,
}
agreeing with~\eyuksing.

\noindent
{\bf Example 2.} As the second example we will consider a model ${\cal M}_2$
for which
we do not know how to construct the mirror manifold ${\cal W}_2$ by means of
the transposition argument~\rGMC.
Let ${\cal M}_2\in\cp{(1,1,3,4,6)}{4}[15]$ be
defined by a transverse polynomial $p_2=0$.
The projectivity condition \eweight\ gives a
$\ZZ_{12}$ fixed point
set and from~(A.3) $k=12$.
$H$ is given by polynomials of order $12$, {\it e.g.} $x_5^2$.
So by setting
$x_5^2={\rm const}$, $x_4^3={\rm const}$ and $x_3^4={\rm const}$ they describe
together with $p_2=x_1^{15} + x_2^{15} + x_3^5 + x_3 x_4^3 + x_3 x_5^2 +
x_2 x_4^2 x_5$ the intersection $H^3$. This system has
$2\cdot 3\cdot 4\cdot 15=360$ solutions. Thus,
\eqn\eXXX{
          H^3~=~360,
}
in agreement with \eyuksing.

In section~5 we return to mirror models of the respective examples where we
 compute the corresponding $(2,1)$-form coupling in the large complex
structure limit. By mirror symmetry the two types of couplings are supposed
to agree. As will be shown this is indeed the case hence verifying mirror
symmetry (at one point in moduli space) for a large class of models.

\newsec{The Fundamental Period}\noindent%Section 3
Recall that the structure of the moduli space of the complex structure is
determined by the period vector $\vp_i$~\refs{\rBG,\rRoll,\rCO}.
Let $\hat p_{\f_j}(y_i)=0$ be a generic hypersurface in
the family $\cp{(k_1,k_2,k_3,k_4,k_5)}{4}[d]$.  Here $\f_j$
denote certain local coordinates on the moduli space.
The periods are integrals of the holomorphic three form $\W(\f_j)$ over a
basis of three cycles $\g_i$, $\vp_i=\int_{\g_i} \W(\f_j)$.
The {\it fundamental period} of
$\W(\f_j)$ is then given by
\eqn\ePER{ \vp_0(\f_j)~~ \define
 ~~ \oint_{B_0} \W(\f_j)~~ = ~~
   {{-}\j d\over(2\p i)^5} \oint_{b_0} {\prod_{i=1}^5 \rd y_i
                               \over \hat p_{\f_j}(y_i)}~.}
where $\j=\f_0$ is up to a constant the coefficient of the term
$\prod_{i=1}^5y_i$ in $\hat p_{\f_j}$.
The integral has been pulled back to the affine space
$\IC^{5}_{(k_1,k_2,k_3,k_4,k_5)}$ and
$b_0=\{y_i|\, |y_i|=\d\,, i=1,\ldots, 5\}$ is determined so as to reproduce
the integral of $\W(\f_j)$ over the fundamental cycle $B_0$.

In this section we will derive an expression for the fundamental  period
associated to a particular one-dimensional deformation
of the mirror manifold $\hat{\cal W}$  of a partial
desingularization $\hat{\cal M}$
of a hypersurface  ${\cal M}\in\cp{(k_1,k_2,k_3,k_4,k_5)}{4}[d]$.
This deformation consists of a sum of five monomials plus a multiple
of a monomial which may be thought of as $\prod_{i=1}^5y_i$.
The derivation will be made twice.
First, the calculation will be carried out
for a class of models in which ${\cal M}$ is defined
by the zero locus of a polynomial $p$, where it is assumed that $p$ is
`invertible'~\rMax. By `invertible' we mean that there exists a
polynomial $p_0$ consisting of five terms such that $p_0$ is transverse,
\ie\ $\rd p_0=p_0=0$ has the origin as its only solution. Then the  fundamental
deformation of ${\cal W}$ is given by adding on multiples of the monomial
$\prod_{i=1}^5 y_i$, where the $y_i$ are
projective coordinates on ${\cal W}$.
After this explicit calculation, the calculation will be redone
in greater generality via toric geometry and Batyrev's proposed construction
of mirror manifolds. In this way, we will be able to
relate the two approaches when they overlap and hence give more evidence
for why the construction discussed in~\rGMC\ gives the mirror partner.

Let ${\cal M}$ be as in section~2, \ie\ a  hypersurface in
$\cp{(k_1,k_2,k_3,k_4,k_5)}{4}$
defined by $p(x_i)=0$ but with $p(x_i)$  an invertible (quasi)homogeneous
polynomial of degree $d$.
The mirror of ${\cal M}$ is an
orbifold, ${\cal W} = \tilde {\cal W}/H$ where $\tilde
 {\cal W}$ is
a hypersurface of degree $\hat d$
defined by $\hat p(y_i)=0$ in $\cp{(\hat k_1,\hat k_2,\hat k_3,
\hat k_4,\hat k_5)}{4}$ obtained by transposing $p$ \rGMC.

We want to calculate the fundamental period for
 the one-dimensional subspace of the complex
structure moduli space of ${\cal W}$
corresponding to deformations along the direction of
$\prod_{i=1}^5 y_i$, \ie\ deformations of $\hat p_0$ given by
\eqn\ePPSI{\hat p_{\j} = \hat p_0 - \j d y_1 y_2 y_3 y_4 y_5.}
There are  several ways of obtaining an explicit expression for
$\vp_0$. On the one hand, following the construction by Candelas {\it et al.}
 \rCdGP,
one can perform the integration explicitly. This is the lead we will
follow. On the other hand \ePER\ satisfies the so-called Picard-Fuchs
equation~\rGriffII.
In the context of string compactifications, the existence of this differential
equation is a general result that follows from the fact that the models we are
considering are $(2,2)$-superconformal string vacua \refs{\rCF,\rLSW,\rCDFLL}.
The Picard-Fuchs equation can be derived from \ePER\ by using Griffiths
`reduction of pole order'  analysis \rGriff. (For examples, see
\refs{\rLSW,\rPicF,\rCF,\rHKTY}.) However, for explicit calculations it is
preferable to use the first method, especially when studying multi-dimensional
moduli spaces.\foot{For $(2,2)$ \LG\ models with $n>5$ superfields it is not
always clear how to perform the generalization of the integral in \ePER~.}
(See also~\rHKTY.)

Leaving the general discussion behind we now go on to compute the integral.
Following \rPeriod\ we expand the denominator around large $\j$ and
evaluate the integral by residues. Leaving the combinatorics to the reader we
straightforwardly obtain the fundamental period as
\eqn\ePERF{\vp_0  = \sum_{m=0}^{\infty}{\G(d\  m+1)\over
\prod_{i=1}^5\G(k_i\,m+1)( d\j)^{d\,m}}.}

The fundamental period  can also be written in terms of a
generalized hypergeometric function,
\eqn\eGHF{\hbox{}\mkern-20mu
%\vp  =
{}_{d{-}1} F_{d{-}2} (\overbrace{{1\over d}{,}\ldots{,} {d{-}1\over d}}{;}
\overbrace{{1\over k_1}{,} \ldots{,}{k_1 {-}1\over k_1}{,}\ldots,{1\over
k_5}{,}\ldots{,} {k_5 {-} 1\over k_5}}{,}1{,}1{,}1{;}
( \prod_{i=1}^5k_i^{k_i}\j)^{-d})}
This form will turn out to be useful when studying the mirror map in the
large complex structure limit.
The overbraces indicate that indices which are common are to be left out. In
general $\vp_0$ is a  ${}_{q} F_{q-1}$ where $q\le (d-1)$. In fact, it is
believed that $q$ is the number of periods which are  related to the polynomial
deformations.

Before generalizing the above via toric geometry,
we review Batyrev's proposed toric construction of mirror pairs
\rBatdual, as enhanced by the discussion of the monomial-divisor mirror map
of Aspinwall-Greene-Morrison \rmdmm.  The discussion will also serve as a
basis for understanding the moduli space as a toric variety.  This is
useful for the computation of the Yukawa coupling in the large complex
structure limit.

Consider an $n$~dimensional integral polyhedron $P$, whose vertices lie in an
$n$~dimensional lattice $M$ in
$M_{\IR}:=M\otimes\IR$.  One associates to $P$ a toric variety $\IP(P)$
\rFul, whose dimension is the same as that of $M$.  The construction also
gives a canonical embedding $\IP(P)\hookrightarrow\IP^{|P|-1}$, where
$|P|$ is the cardinality of $P\cap M$, and the coordinates of $\IP^{|P|-1}$
are identified with the points of $P\cap M$.  Consider the algebraic torus
$T=\Hom(M,\IC^*)\simeq(\IC^*)^n$, and identify $M$ with the lattice of
characters $\Hom(T,\IC^*)$
of $T$.  $T$ may be embedded in $\IP^{|P|-1}$ via the map defined by
$t\mapsto (m_1(t),\ldots,m_{|P|}(t))$, where the $m_i$ range over the
points of $P\cap M$, identified as characters (sometimes called monomials)
of $T$.  The toric
variety $\IP(P)$ is in fact the closure of this map.  We will often think of
$\IP(P)$ as a projective variety in this fashion without further comment.

A {\it reflexive} polyhedron is an integral polyhedron containing 0 in its
interior, such that each facet of $P$ (that is, a codimension 1 face of $P$)
is supported by a hyperplane $H$ which can be defined by a linear equation
of the form $H=\{\,y\in M_{\IR}\mid \langle\ell,y\rangle =-1\,\}$ for some
$\ell$ in $N:=\Hom(M,\ZZ)$. Note that $N$ and $M$ are dual lattices,
and $N$ is canonically identified with the lattice $\Hom(\IC^*,T)$
of one parameter subgroups of $T$.
Batyrev shows that if $P$ is reflexive, then
the general hyperplane section of $\IP(P)$ is \CY\ (possibly with mild
singularities, which can be resolved to obtain a \CY\ manifold).  Put
$N_{\IR}:=N\otimes\IR$.  The {\it polar polyhedron} (which Batyrev calls the
dual polyhedron) is given by
\eqn\polar{P^\circ=\{\,x\in N_{\IR}\mid\langle x,y\rangle\ge -1\
\hbox{for all } y\in P\,\},}
and is reflexive if and only if $P$ is reflexive.  Batyrev proposes that the
hyperplane sections $\Mbar$ of $\IP(P)$ and $\Wbar$
of $\IP(P^\circ)$ should form a
mirror pair.

Consider the weighted projective space $\wp$,
and let $d=\sum_i k_i$.  The study of hypersurfaces of degree $d$ in $\wp$
is related to the study of hyperplane sections of the variety obtained
as the image of the mapping $f$ from $\wp$ to $\IP^{|P|-1}$ defined by
$f(x)= (m_i(x))$, as $m_i$ ranges over all ($|P|$)
degree $d$ monomials in $\wp$.  The mapping $f$ need not be defined
everywhere, so that
it is only a rational mapping. Let $V$ be the closure of the image of $f$.
But from the above discussion, we have a nice
description of $V$.  It is natural to define $P$ as the convex hull of
the set of exponents of all degree $d$ monomials in $x_1,\ldots,x_5$.
To make contact with the definition of reflexivity given above, we will need to
translate $P$ by the vector $(-1,-1,-1,-1,-1)$.  The translate is needed to
make the origin a point of $P$.  So let
\eqn\subl{M=\{\,x\in\ZZ^5 \mid \sum k_ix_i=0\,\},}
where the restriction is equivalent to the \CY\ condition
$\sum_{i=1}^5k_i=d$.
 In terms of $M$, which is a rank 4 lattice, we define
\eqn\poly{P=\hbox{the convex hull of }\{\,x\in M\mid x_i\ge -1\
\forall i\,\}.}
It is then clear that $V=\IP(P)$. When the weights $k_i$ need emphasis, we
will write $P_{\vec k}$ in place of $P$, where ${\vec k}$ is short for
$(k_1,\ldots,k_5)$.

Since we want our hyperplane sections to have \CY\ resolutions,
we must now make the
assumption on $(k_1,\ldots,k_5)$ that $P_{\vec k}$ is reflexive.
In particular, $\dim P_{\vec k}=4$, which leads quickly to the conclusion
that $V$ is birational to $\wp$.
The classification of reflexive polyhedra in dimension~4 has not been done.
However, there is a result which shows that there are many
examples of $\vec k$ for which $P_{\vec k}$ is reflexive.  Before stating it,
recall the map  $\r:\CP{4}\to\cp{(k_1,k_2,k_3,k_4,k_5)}{4}$
defined by $\r(x_1,\ldots,x_5)=(x_1^{k_1},\ldots,x_5^{k_5})$.  Let
$\a_j=e^{2{\p}i/{k_j}}$.  Then $\r$ exhibits $\cp{(k_1,k_2,k_3,k_4,k_5)}{4}$
as an orbifold of $\CP{4}$
by the
group $\ZZ_{k_1}\times\cdots\times\ZZ_{k_5}$, since $\r$ respects the
automorphism
$(x_1,\ldots,x_5)\mapsto(\a^{r_1}x_1,\ldots,\a^{r_5}x_5)$ of $\CP{4}$ for any
integers $r_i$.

Let $\cM\subset\wp$ be a degree $d$ hypersurface, and
let $X=\r^{-1}({\cM})\subset\CP{4}$.
We say that $\cM$ is {\it quasismooth} if $X$ is smooth.  Then the following
is shown in \rCOK.

\bigskip\noindent
{\bf Proposition.} Suppose there is a quasismooth degree $d$ hypersurface
$\cM\subset\wp$.  Then $P_{\vec k}$ is reflexive.

\bigskip
These weighted projective spaces have been classified \rAR.  There
are 7555 of them.  In addition, there are many other examples of $\wp$ for
which $P=P_{\vec k}$ is reflexive \rCOK.

There are 5 coordinate functions on the $\ZZ^5$ which
contains $M$ according to~\subl; restricting to $M$, these are naturally
thought of as elements $v_1,\ldots,v_5$ of $N$, where $v_i(x)=x_i$ for
$x\in M$.
{}From the definitions \polar\ and \poly,
it follows immediately that $P^\circ$ contains the six
points $${\vec 0},v_1,v_2,v_3,v_4,v_5.$$
To these six points are associated respective linear coordinate functions
$m_i$ on $\IP(P^\circ)$, with $0\le i\le 5$, coming from the projective
embedding of $\IP(P^\circ)$ and our earlier discussion.

The $v_i$ are
linearly dependent; the only relation between them (up to scalar) is
$\sum_ik_iv_i=0$.  This leads to the relation
\eqn\erelate{
\prod_{i=1}^5 m_i^{k_i}=m_0^d}
between the $m_i$.
It is easy to see that all relations between the $m_i$ are just powers of
this one.

Choose the one parameter family of hypersurfaces
\eqn\etorfam{
\sum_{i=1}^5m_i-\psi dm_0=0~.}
In \rBatyrev, a natural period is calculated for toric hypersurfaces; when
this result is applied to the family \etorfam, the
fundamental period \ePERF\ is once again obtained.
Since these two techniques are
rather different, it is reassuring that the results agree.

In fact \etorfam\ and the family it describes can be related to
the one parameter family \ePPSI\ in
$\cp{(\hat k_1,\hat k_2,\hat k_3,
\hat k_4,\hat k_5)}{4}[\hat d]$  by a fractional transformation.
 However, there are identifications due to the non-linear change
of variables. These are precisely described by performing a quotient by  $H$.
Thus, in those cases when the transposition scheme can be performed
the result is recovered by the more general toric construction of Batyrev,
and the families \ePPSI\ and \etorfam\ are identified
\refs{\rCOK,\rBK}.  Not surprisingly, it can be shown that the
integration cycle used in \rBatyrev\  coincides with the integration cycle
considered in \ePER\ after the identifications are made.

\noindent
{\bf Example}: Consider a one parameter family of deformations
${\cal W}=\tilde {\cal W}/G$ where $\tilde {\cal W}
 \in \cp{(1,3,3,3,5)}{4}[15]$ is defined by
\eqn\edefex{
\hat p_\j=\hat p_0 - 10 \j  \prod_{i=1}^5 y_i~
=y_1^{10}y_5+y_5^3+y_2^5+y_3^5+y_4^5 - 10 \j \prod_{i=1}^5 y_i,}
and $G=(\ZZ_5:4,1,0,0,0)\times (\ZZ_5:4,0,1,0,0)$.\foot {We use
the notation $(\ZZ_r:\Q_1,\Q_2,\Q_3,\Q_4,\Q_5)$ for a
$\ZZ_r$ symmetry
with the action $(y_1,y_2,y_3,y_4,y_5) \to (\a^{\Q_1}
y_1,\ldots,\a^{\Q_5} y_5)$, where
$\a^r~=~1$.}
The peculiar form of $\hat p_0$ is due to it being the transpose of
$p_0=x_1^{10}+x_1x_5^3 + x_2^5 + x_3^5 + x_4^5$, where $p_0=0$
is the mirror ${\cM}\in\cp{(1,2,2,2,3)}{4}[10]$ of ${\cal W}$.
We want to relate~\edefex\
to the corresponding realization of ${\cal W}$ as a hypersurface in a toric
variety given by~\etorfam. To do so  we make the identifications
\eqn\eident{
m_0=y_1y_2y_3y_4y_5\,,\quad m_1=y_1^{10}y_5\,,\quad m_i=y_i^5\,,i=2,3,4
\,,\quad m_5=y_5^3~,}
where $m_1m_2^2m_3^2m_4^2m_5^3=m_0^{10}$
 from~\erelate. This map is not well defined since
 the $m_i$ are invariant under
a $\ZZ_3\times\ZZ_5^3$ action generated by
$G\times(\ZZ_{15}:1,3,3,3,5)$ on the $y_i$; $m_0$ transforms under a
$\ZZ_{10}$ under the group of rescalings of the $y_i$ which preserve
the $m_i$ for $i\neq 0$.
Thus, to make the identification $1-1$ we have to consider a quotient
by $G\times\ZZ_{15}$. But the $\ZZ_{15}$ is already enforced by the
projectivization in $\cp{(1,3,3,3,5)}{4}[15]$. Hence, we are left with
$\cp{(1,3,3,3,5)}{4}[15]/G$ which is the result obtained using
the transposition argument.

\newsec{The Moduli Space of The Mirror}%Section 4
\noindent
With the fundamental period at hand we turn to its behavior around the
singular points in the moduli space. The computation is by now standard,
follows that of~\rCdGP\ and can be found in appendix~B. Suffice it to say that
we in this way understand the monodromy around $\j=\infty$.

Our next step is to focus on the
 complex structure moduli space, and in particular the
large complex structure and $\j=\infty$ limit points.
We begin by reviewing the set-up leading to the
monomial-divisor mirror map from \rmdmm.

In the previous section the toric variety $V=\IP(P)$ was described as
embedded in $\IP^{|P|-1}$.
 $V$ can also be constructed from the {\it normal
fan} $\Np$ of $P$; in this situation, this is the fan in $N_{\IR}$ obtained by
taking the union of the cones over all proper faces of $P^\circ$. To
begin to resolve the  singularities of $V$, we can take a simplicial
subdivision of $\Np$ to arrive at a new fan $\D$.  It is shown in \rOP\ that
one can choose $\D$ in such a way that the resulting toric variety $\hat{V}$
(which
is an orbifold since the subdivision is simplicial) is projective, and
that the edges of the fan are precisely the edges spanned by the points of
$(P^\circ\cap N)-\{0\}$.

A key idea leading to the monomial-divisor mirror map is that the points of
$P^\circ\cap N$ have two interpretations.  We have already discussed its
interpretation as monomials on $\IP(P^\circ)$ (since $N$ plays the role for
$P^\circ$ that $M$ did for $P$).  On the other hand, points of $P^\circ\cap N
-\{0\}$ span edges of $\D$, and as such give rise to toric divisors on
$\hat{V}$.

A special role is played by the points of $P^\circ\cap N$ which are in the
interior of a facet.  The corresponding divisors of $\hat{V}$ are exceptional
for the natural map $\hat{V}\to V$ and have image equal to a point; hence
they are disjoint from the proper transform $\Mhat$ of a general hyperplane
section $\Mbar$ of $V$.  Let $(P^\circ\cap N)_0$ denote the points of
$\PN$ which do not lie in the interior of a facet.

Now, similarly choose a subdivision $\D^\circ$ of ${\cal N}(P^\circ)$,
getting 4~dimensional toric varieties $\hat{U}\to U$ and hypersurfaces
$\hat{\cal W}\to
\bar{\cal W}$ analogous to $\hat{V},\ V,\ \Mhat, \hbox{and }\Mbar$.

The monomial-divisor mirror map is the natural isomorphism
\eqn\emdmm{
H^{2,1}_{\scriptstyle {\rm poly}}(\hat{\cal W})\simeq
((\ZZ^{(\PN)_0-\{0\}})/M)\otimes\IC
\simeq H^{1,1}_{\scriptstyle{\rm toric}}(\Mhat).}
The subscripts ``toric'' and ``poly'' refer to the subspaces generated
by polynomial deformations of $\hat{W}$ and toric divisors of $\Mhat$,
respectively.  See \rmdmm\ for more details.

The next step is to examine the moduli space of the mirror.  According to
\rmdmm, we let $N^+=N\oplus \ZZ$, and lift the set $(\PN)_0$ to the subset
of the affine hyperplane $\{(n,1)\}\subset N^+$ which projects to it.  We
form the secondary fan \rOP\ associated with this set of points.
This fan is a finite rational
polyhedral subdivision of $H^{1,1}_{\scriptstyle {\rm toric}}({\Mhat})$.
The toric variety associated with the secondary fan is to be viewed as a
``simplified moduli space'' for $\hat{W}$, describing all hypersurfaces
of the form $\sum c_im_i$, where the $m_i$ range over the monomials
corresponding to points of $(\PN)_0$.  There is a natural map to the polynomial
part of the true moduli space.  Assuming that this map is dominant,
i.e.\ has dense image,
a precise form of mirror symmetry is conjectured in \rmdmm.  We will assume
this conjecture to be true, and will use it without verifying dominance.
Since our conclusions are consistent with algebraic geometry, we can view
this as providing evidence for the conjectured form of mirror symmetry, as
well as for dominance.

We now need to interpret this moduli space using the secondary fan.  By general
toric principles~\rFul,
$s$~dimensional cones of the secondary fan correspond to
codimension~$s$ toric subvarieties of the simplified moduli space.

Note that our one parameter family is determined inside the simplified
moduli space by setting the coefficients
of all monomials to 0, except the monomials
$m_0,\ldots,m_5$.  Our one~parameter family is clearly invariant
under the torus
(that is,
under rescalings of the coefficients), hence corresponds to a codimension~1
cone in the secondary fan.  It follows immediately from the standard toric
correspondence (see section~3.1 of \rFul)
that this face is the one spanned by the classes in $H^{1,1}$
of all the exceptional
divisors of $\hat{\cal M}$.

Since the orbifold K\"ahler cone of $\hat{\cal M}$ appears in the secondary
fan \rmdmm\ and the hyperplane class $H$
is an edge of the K\"ahler cone, it follows that $H$ is an edge of
the secondary fan.  Note: different orbifold
K\"ahler cones can arise from different
choices of $\D$; but $H$ is an edge of any of these.  This is consistent with
the discussion following \eYUK.

The toric correspondence between edges and divisors (p.~60 of \rFul)
gives a divisor $D_H$ in the simplified
moduli space.  $D_H$ passes through all large complex structure
limit points.
$D_H$ also meets the one~parameter family at the point $\psi=\infty$, a point
which is invariant under the torus.  This point of intersection
corresponds (via section~3.1 of \rFul)
to the cone spanned by the class of $H$ and those of the
exceptional divisors.  Denote this cone by $\sigma$.

Suppose that the cardinality of $(\PN)_0$ is $n$.
Then the secondary fan
sits inside $\IR^{n-5}$.  Ignoring the points $\vec{0},v_1,\ldots,v_5$,
we see that there are $n-6$ exceptional divisors $E_i$.
Let $L\subset H^{1,1}_{\scriptstyle{\rm toric}}({\cal M},\ZZ)$ be the lattice
spanned by $H$ and the $E_i$.  As before, let $k$ denote the weight of
the hyperplane class of ${\cal M}$.

{\bf Claim}. $L$ is a sublattice of
$H^{1,1}_{\scriptstyle{\rm toric}}(\Mhat,\ZZ)$ of index $k$, hence a
neighborhood of the point $\j=\infty$ in the simplified moduli space is
given as the quotient of $\IC^{n-5}$ by a group of order $k$, as described
in section~2.2 of \rFul.

We will see presently that more is true: the group is in fact $\ZZ_k$.

To see this, we perform the calculation in a convenient choice of coordinates
on $H^{1,1}_{\scriptstyle{\rm toric}}(\Mhat,\ZZ)$.  Note that elements of
$\Hom(H^{1,1}_{\scriptstyle{\rm toric}}(\Mhat,\ZZ),\ZZ)$ may be thought of
as linear dependencies between the points
of $(\PN)_0$, identified as a subset of $N^+$ as discussed earlier in this
section.  In particular, $\vec{0}\in(\PN)_0$ also corresponds to a
divisor class $D_0$ (but not an effective divisor).  Now consider an
exceptional divisor $E\subset\Mhat$.  Since the $v_i$ span $N$,
the vector $v\in(\PN)_0$ corresponding to $E$ can be expressed as an integral
linear combination of the $v_i$. Using a superscript of ``$\scriptstyle{+}$''
to denote the lifting of a vector to $N^+$, there is still
a similar expression expressing $v^+$ as an integral linear combination
of $\vec{0}^+$ together with the $v_i^+$.  In addition, the relation
$\sum_ik_iv_i=0$ gives rise to the relation $\sum_ik_iv_i^+-d\vec{0}^+=0$.
These linear dependencies clearly
span all linear dependencies, and so give rise to coordinates on
$H^{1,1}_{\scriptstyle {\rm toric}}(\Mhat,\ZZ)$.  Note from the choice of
linear dependencies that the exceptional divisors
$E_i$ have as coordinates the standard unit
vectors $e_i$ for $1\le i\le n-6$, while $D_i$ has last coordinate $k_i$.
By writing $H=\sum_{i=1}^5 b_iD_i+\sum c_iE_i$ and comparing
degrees in $\wp$, we see that $k=\sum b_ik_i$. On the other hand the last
coordinate of the right hand side is $\sum b_ik_i+\sum c_i\cdot 0=\sum b_ik_i$.
Hence $H$ has last coordinate $k$.  This immediately justifies the claim.

For emphasis, we include a quick example.

\noindent
{\bf Example:} Let us consider
$H^{1,1}_{\scriptstyle {\rm toric}}(\Mhat,\ZZ)$ for
${\cal M}\in \cp{(1,2,2,2,3)}{4}[10]$.
Here we have divisors $D_1,\ldots,D_5$ corresponding to the proper transforms
of $x_i=0$, and two exceptional divisors.  It is clear from our definitions
that in this case (or any example with $k_1=1$), the vectors
$v_2,v_3,v_4,v_5$ form a $\ZZ-basis$ for $N$.
In our choices of coordinates we have the following:

$$\matrix{D_1 & (-2,-2,-2,-3,1) & 0 & 0 & 1 \cr
          D_2 & (1,0,0,0,1)     & 1 & 1 & 2 \cr
          D_3 & (0,1,0,0,1)     & 1 & 1 & 2 \cr
          D_4 & (0,0,1,0,1)     & 1 & 1 & 2 \cr
          D_5 & (0,0,0,1,1)     & 2 & 1 & 3 \cr
          E_1 & (-1,-1,-1,-2,1) & 1 & 0 & 0 \cr
          E_2 & (-1,-1,-1,-1,1) & 0 & 1 & 0 \cr
          D_0 & (0,0,0,0,1)     &-6 &-5 &-10}.$$
The meaning of the above table is that each of the last three columns give
coefficients of linear dependencies among the coordinates in the second column.
The vectors in the second column are written in the choice of basis for
$N_{\IR}$ just mentioned, then lifted to $N^+$ by appending a ``1''.
Note that the entry in the last column after the
vector corresponding to $D_i$ is $k_i$; this is a general fact, arising
from the relation $\sum k_iv_i^+-d\vec{0}^+=0$, as mentioned above.
The last 3~columns give the coordinates of the divisor class represented by
the divisor in the first column.
For this example, we calculate $k=6$; the hyperplane class turns out
to have coordinates $(4,3,6)$.
We will return to this example when we discuss monodromy.

\newsec{The Yukawa Coupling}
\noindent%Section 5
We start by studying the monodromy around the large complex structure point.
Knowing this will enable us to choose the relevant flat coordinate, $t$
for which the Yukawa coupling in the $t$-coordinate will coincide with
the intersection number computed in section~2.

\subsec{Monodromy and The Large Complex Structure Limit}\noindent
Recall the discussion from the last section.
We have a part of the moduli space described by a cone $\sigma$ with edges
corresponding to the exceptional divisors and to $H$. In coordinates, these
are the standard unit vectors $e_i$ for $1\le i\le n-6$ and a vector whose
last coordinate is $k$.
It now follows from toric generalities \rFul\
that the point of the toric moduli space corresponding
to $\sigma$ is singular if $k>1$; in fact, the point is locally the
quotient of a polydisc by the group $\ZZ_k$.  We will illustrate this
by an example presently.
Letting $T_\infty$ denote
the monodromy about this point within the one parameter family, it follows that
the monodromy about a general point of $D_H$ is $T_\infty^k$.  To see this,
standard toric techniques and the above description of the generators of
$\sigma$ show that the simplified moduli space is locally
an orbifold of the form $\IC^{n-5}/\ZZ_k$, where $\ZZ$ acts by
$(x)\mapsto (\zeta^{a_i}x_i)_i$, where $\zeta=e^{2\pi i/k}$ and $a_{n-5}=1$,
while $D_H$ is given in these coordinates by $x_{n-5}=0$ and the one~parameter
family is given by $x_1=\cdots =x_{n-6}=0$.  The loop
$(p_1,\ldots,p_{n-6},e^{2\pi i\theta})$ where the $p_i$ are fixed
and non-zero, while $\theta$
ranges from 0 to 1 is seen to go around $D_H$ once; but as we let the $p_i$
approach 0, we see that this loops around $D_H$ a total of $k$ times.  This
gives the claimed relation between the various monodromies up to conjugacy,
which will not affect the calculation of the Yukawa couplings in the
sequel.  This will also be illustrated by the following example.

\noindent
{\bf Example.}  Let us now continue our discussion of
a hypersurface ${\cal M}\in\cp{(1,2,2,2,3)}{4}[10]$.
We have seen that the toric moduli space is describe by the cone $\sigma$
with edges
\eqn\edges{(1,0,0),\ (0,1,0),\ (4,3,6)~.}
(The first two coordinates of the last edge have not been explained here; this
is safely done, since the result does not depend on these coordinates.)
The toric variety for $\IC^3$ is described by the cone with edges
\eqn\cedges{(1,0,0),\ (0,1,0),\ (0,0,1)}
and there is an obvious linear transformation ${\cal L}$ mapping the
second cone to
the first, identifying the first two edges of the first cone with the
first two edges of the second cone.  But the toric varieties are different,
since the integral lattices have changed (the linear transformation used
has determinant~6).  In other words, the map of tori is not an isomorphism.
We are using the standard torus $(\IC^*)^3\subset\IC^3$
in identifying the second toric variety with $\IC^3$ in the usual way.
The induced map $\nu$ on the toric varieties, when restricted to the torus,
is just
\eqn\quomap{\nu(x_1,x_2,x_3)=(x_1x_3^4,x_2x_3^3,x_3^6).}
This follows because the matrix of ${\cal L}$ gives rise to a map between the
spaces of one parameter subgroups associated to the toric varieties; we
must transpose this matrix to get the map between the associated spaces of
characters, and this immediately gives \quomap.
The group we must quotient out by is evidently $\nu^{-1}(1,1,1)$, which is
the group $(\ZZ_6:2,3,1)$.

In the $(x_1,x_2,x_3)$ coordinates just introduced, the divisor $D_H$
corresponds
to $x_3=0$.  But $x_3$ is only a multi-valued function on the moduli space.
Near a general point of $D_H$, we have that $x_1$ and $x_2$ are non-zero; so
$D_H$ has ($\ZZ_6$~invariant) equation $x_1x_2x_3=0$.  Now look at the loop
$\theta\mapsto(x_1,x_2,e^{2\pi i\theta})$.  This clearly loops around
$x_1x_2x_3=0$
once.  But as $x_1$ and $x_2$ approach 0, we must take $x_3^6$ as our invariant
equation; that is, the loop winds around 6~times, illustrating the general
situation.

We now know that $T^k_{\infty}$ gives
the monodromy around the large complex structure limit point.
On general grounds the Picard-Fuchs equation satisfied by $\vp_0$ has at
$\j=\infty$ a quasi-unipotent point of index $l$ \rGriffII.
By this we mean
that there exists a natural number  $k$, defined as above, such
that
\eqn\eunipotent{
U^{l-1}~\neq~0~,\qquad U^l~=~0~,\qquad {\rm where}~~ U~=~ T_{\infty}^k - I~.}
It was observed that
in fact the Yukawa coupling in the large complex structure limit can be read
off from $U^3$ \rCdFKM\ and hence that $l=4$.\foot{Thus, based on our
explicit calculation, discussed below, we conjecture that any Picard-Fuchs
equation with solutions $\vp_i$ as given by~(B.7) have a quasi-unipotent
point, $\j=\infty$, of index $4$.}
 In an integral and symplectic
basis one can show that $U^3=y^{\scriptstyle {\rm lcs}}_{\j\j\j}E$
where $E$ is the  zero-matrix except
for one entry along the first column which is $1$ \rCdFKM. In our basis, $U^3$
has more than one non-vanishing entry in the first column.
However, by using that
we know two of the basis elements in the integral basis one can  show that the
smallest (in terms of absolute value) of the non-zero entries in $U^3$
gives the Yukawa coupling, $y_{\j\j\j}^{\scriptstyle {\rm lcs}}$.

Thus, from \eunipotent\ it follows that the Yukawa coupling is
readily computed for all of the 7,555 transverse hypersurfaces in weighted
$\CP{4}$. All we have to do is to calculate $U^3$. However, there is a slight
computational problem in that there are models for which
$b_3\sim O(10^3)$~\rAR, \ie\
the number of periods is $O(10^3)$. Combined with the
fact that for some of these models $k\sim O(10^8)$, merely computing
$T_{\infty}^k$ would take quite some time! Fortunately, the problem is very
much simplified because of the special form that the matrices $T$ and $A$,
and hence $T_{\infty}$, take, see~(B.17) and
table~B.1. The first step is to put
$T_{\infty}$ on a  block-diagonal form. This not only makes the computation
feasible in real time but also shows that the Yukawa coupling can be written
as
\eqn\eblock{
y^{\scriptstyle {\rm lcs}}_{\j\j\j}~=~k^3 {C_1\over C_2}~,
}
where the $C_i$ can be found in terms of the basis which (block) diagonalizes
$U$.\foot{In the next section we will show how to compute $C_1/C_2$
independently for all models.}
 It turns out that, in this particular situation, to compute $C_i$ is an
$O(N^2)$ process for an $N\times N$ matrix while matrix multiplication is
$O(N^3)$. This is what saves the day and makes it possible to put the problem
on a computer. The calculation has been performed for some 7,400 models. For
the remaining ones it  has not yet been possible  to construct the matrix $A$
associated to the phase symmetry ${\cal A}$. However, we believe that this
is merely a technical problem and that within the near future all couplings
will  be found by this method.

\subsec{The Flat Coordinate and The Yukawa Coupling}\noindent
We now turn to studying the flat coordinate, $t$ which will
provide the mirror map between ${\cal M}$ and ${\cal W}$.
We then go on to  give the general expression for the Yukawa coupling
$y_{\j\j\j}$ as a function of $\j$ for ${\cal W}$.  Knowledge of $t$, and hence
the mirror map, enable us to map $y_{\j\j\j}$ to $y_{ttt}$ where $y_{ttt}$ is
the  Yukawa coupling
on ${\cal M}$ corresponding to the K\"ahler deformation.

The na\"\ii ve flat coordinate, relevant for the $\j\to\infty$ limit is defined
by~\rCdGP\
\eqn\eTflatdef{T_{\infty}~:~\tilde t\to \tilde t+1~.}
However, from the previous discussion we know that the monodromy
around the large complex structure limit point is given by $T_{\infty}^k$.
Hence, the natural flat coordinate from that point of view is
\eqn\eTflatdefII{T_{\infty}^k~:~t\to t+1\,,\qquad t=\tilde t/k~.}
In fact $t=\tilde t/k$ is the analog of $\r^*(H)=k\tilde{H}$ where
$\tilde H$ is the hyperplane class induced from $\IP^4$ and
$H\in H^2({\cal M},\ZZ)$ as discussed in appendix~A.
{}From \eTflatdef\ and \eTflatdefII\ we can find $t$ as a linear
combination of the periods
$\vp_j$ given by (B.7). By
analytically continuing the $\vp_j$ to large $\j$, $t(\j)$ will give the
mirror map. Although this procedure can straightforwardly be carried out along
the lines of~\rCdGP\ we are only interested in the large $\j$ behavior of
$t(\j)$. This is obtained by studying  the Riemann ${\cal P}$-symbol
associated to the generalized hypergeometric equation to which the $\vp_j$
are the solutions. In fact by looking at~\eGHF\ we readily read off
that among the solutions as $\j\to\infty$ there are four of the type
$(1,\log((\j d)^d),\log^2((\j d)^d),\log^3((\j d)^d))$. Thus, using this
and~\eTflatdefII,
\eqn\eTFLATA{
t\sim -{1\over k}{d\over 2\p i}\log(\j d)\,\qquad \j\to\infty~.}
Since $t$ is the coordinate on the K\"ahler moduli space and $\j$
parametrizes the fundamental deformation on the complex structure,
eq.~\eTFLATA\ gives the mirror map for large $\j$.

In \refs{\rRoll,\rCdGP} it was shown that the Yukawa coupling on the space of
complex
structure deformations is given by
\eqn\eYUK{
\eqalign{y_{\j\j\j}& =\int \W \wedge {\del^3 \W\over \del\j^3} \cr
\phantom{y_{\j\j\j}}& = z^a \left({d\over d\j}\right)^3 {\cal G}_a -
                       {\cal G}_a \left({d\over d\j}\right)^3 z^a~,}
}
where we restrict to the fundamental  deformation. The $({\cal G}_a,z^b)$
are the components of an integral and symplectic basis. In particular, ${\cal
G}_0=\vp_0$ and $z^0=\vp_0-\vp_1$~\rCdGP. Knowing these two basis vectors
is what makes it possible to make a (partial) change of basis from the $\vp_j$
to the $(z^a,{\cal G}_a)$ and hence to read off
$y_{\j\j\j}^{\scriptstyle {\rm lcs}}$ from $U^3$ as discussed
in the previous subsection.
 Unfortunately, it does not seem to be
possible to construct the complete symplectic basis with only
the knowledge of the
conifold singularity.

The Yukawa couplings can also be found  by studying the chiral ring of
the underlying
$N=2$ superconformal field theory~\rLeVaWa. The chiral ring is
equivalent to the ring  of monomials based on the defining equation, $\hat
p_\j$, of ${\cal W}$ modulo the ideal generated by $\rd\,\hat p_\j$. On general
grounds it has been shown that the unnormalized coupling is given by \rVTop\
 \eqn\eHes{y_{\j\j\j} = {(\prod_{i=1}^5 y_i)^3\over {\cal H}}~,}
where ${\cal H}$, the Hessian, is the determinant of the matrix of
second derivatives
of $\hat p_\j$. In general, one would have to use the  ring to rewrite
$(\prod_{i=1}^5 y_i)^3$ to a monomial proportional to ${\cal H}$. This would
necessarily involve knowing the explicit form of $\hat p_\j$. However, we
will  argue  that one can find $y_{\j\j\j}$ without explicitly referring to the
$\hat p_\j$. Recall that in general there does not exist a representation of
${\cal W}$ as a hypersurface in a weighted projective space although
${\cal W}$ is always well defined as a hypersurface in a toric variety.

The first point  to observe is that because of the conifold singularity at
$\j=(\prod_{i=1}^5k_i^{k_i/d})^{-1}$ the coupling will also have such a
singularity. This can be understood by looking at eq.~\eYUK\ . The periods in
the symplectic basis are linear combinations of the $\vp_j$ which have
conifold  singularities. Hence, the denominator in \eHes\ must have the form
$(1-\prod_{i=1}^5k_i^{k_i}\j^d)$, up to an overall numerical factor.

  The next point is to use   that we know  $y_{\j\j\j}$  in the
large complex structure limit from the monodromy calculation
in the previous section. In terms of the flat coordinate and the large
complex structure gauge \rCdGP\ the coupling is given by
\eqn\eYTTT{
y_{ttt}=  {y_{\j\j\j}\over \vp_0^2} \left({d\j
\over dt}\right)^3~.
}
{}From \eTFLATA\  we have
 $\left({d\j\over dt}\right)\sim -k{2\p i\over d} \j$. Thus,
in order for  \eYTTT\ to be consistent we find that
\eqn\eYJJJ{
y_{\j\j\j}=\left({d\over 2 \p i}\right)^3
{y^{\scriptstyle {\rm lcs}}_{\j\j\j}\over k^3}
{\prod_{i=1}^5 k_i^{k_i}\j^{d-3}
\over (1-\prod_{i=1}^5k_i^{k_i}\j^d)}~.}
Because of the mirror map, \eTFLATA\ and \eYTTT\ and
by expanding~\eYJJJ\ around large $\j$,
$y_{\j\j\j}^{\scriptstyle {\rm lcs}}$ is predicted to give
the $(1,1)$ form Yukawa coupling in the large radius limit on ${\cal M}$, \ie\
the intersection number $y_{JJJ}$ calculated in section~2.
Note that in terms of the flat coordinate $\tilde t$, the Yukawa coupling as
$\j\to\infty$ is $y_{\j\j\j}^{\scriptstyle {\rm lcs}}/k^3$. This is the lowest
order term in the instanton expansion as computed for some
examples in~\rCdFKM.

Let us now show how we in fact can compute
$y_{\j\j\j}^{\scriptstyle {\rm lcs}}$ analytically. From~\rCdGP\ an integral
symplectic basis can be chosen such that under monodromy around the
conifold, $({\cal G}_i,z^i)\to({\cal G}_i+\d_{i,0}z^0,z^i)$. Thus, only
${\cal G}_0$ diverges logarithmically as we approach the
singularity~(B.14).\foot{To be more precise,
one has to study the Riemann ${\cal P}$-symbol associated to the
differential equation to which the periods are solutions. One then finds that
\hbox{$\vp_j\sim(\j-(\prod_{i=1}^5k_i^{k_i/d})^{-1})^{l_j}$}, $l_j\in\BM{N}$
for all periods except one,
${\cal G}_0\sim z^0\log(\j-(\prod_{i=1}^5k_i^{k_i/d})^{-1})$ with
$z^0\sim (\j-(\prod_{i=1}^5k_i^{k_i/d})^{-1})$ as
$\j\to(\prod_{i=1}^5k_i^{k_i/d})^{-1}$.}
It then follows that in computing $y_{\j\j\j}$ from~\eYUK\ in the
limit $\j\to(\prod_{i=1}^5k_i^{k_i/d})^{-1}$ the leading term
comes from $z^0\left({d\over d\j}\right)^3 {\cal G}_0$. From~(B.14) the
$\vp_i$ are appropriate for studying the conifold singularity and hence it is a
straightforward, though a bit tedious, exercise to obtain $y_{\j\j\j}$ as
$\j\to(\prod_{i=1}^5k_i^{k_i/d})^{-1}$. We find, using~\eYUK, that
the leading behavior of $y_{\j\j\j}$ as we approach the conifold singularity
is given by
\eqn\elimit{
%\lim_{\j\to(\prod_{i=1}^5k_i^{k_i/d})^{-1}}
y_{\j\j\j}\sim
\left({d\over 2\p i}\right)^3{d\over \prod_{i=1}^5 k_i}
\left(\prod_{i=1}^5k_i^{k_i/d}\right)^3
\sum_N\left(\j\prod_{i=1}^5k_i^{k_i/d}\right)^{dN}\,.
}
Comparing with~\eYJJJ\ by taking the conifold limit
%
%\eqn\elim{
%\left({d\over 2 \p i}\right)^3
%{y^{\scriptstyle {\rm lcs}}_{\j\j\j}\over k^3}
%{\prod_{i=1}^5 k_i^{k_i}\j^{d-3}
%\over (1-\prod_{i=1}^5k_i^{k_i}\j^d)}\to
%\left({d\over 2\p i}\right)^3 {y_{\j\j\j}^{\scriptstyle {\rm lcs}}\over k^3}
%\left(\prod_{i=1}^5k_i^{k_i/d}\right)^3
%\sum_N\left(\j\prod_{i=1}^5k_i^{k_i/d}\right)^{dN}\,,
%}
%
we readily read off
\eqn\eYES{
y_{\j\j\j}^{\scriptstyle {\rm lcs}}~=~{d\, k^3\over \prod_{i=1}^5 k_i}\, .
}
Thus, we get  perfect agreement with~\eyuksing!

To illustrate the ideas explained above let us now return to the
 examples from section~2.

\noindent
{\bf Example 1.} Consider the hypersurface
${\cal M}_1\in\cp{(1,1,1,2,2)}{4}[7]^{2,95}_{-186}$ defined
by $p_1=0$ where $p_1$ is a transverse polynomial. (As usual the sub- and
superscripts refer to $\EU$ and $b_{1,1},\,b_{2,1}$ respectively on a
\CY\ desingularization of ${\cal M}_1$.)
{}From \ePERF\ the period
corresponding to the fundamental deformation $\prod_{i=1}^5y_i$ on the mirror
 manifold ${\cal W}_1$ is
\eqn\eEXONE{\vp_0 ~=~\sum_{m=0}^{\infty}{\G(7m+1)\over \G^2(2m+1)
 \G^3(m+1) (7 \j)^{ 7m}}.}
 From~(B.16) we have $c_j=(1,1,-2,-1,4,-1,-2)$. The number of linearly
 independent periods is $7-1=6$ since there is only one relation among the
$\vp_j$.\foot{Note that on ${\cal W}_1$, $2 b_{2,1} + 2=6$ and we get
all of the periods through
the action of the phase symmetry ${\cal A}$.
In general this will not be the
case. The missing periods are the complex deformations which cannot be
written as  polynomial deformations~\rGH.}
{}From
$$
\sum_{j=0}^6 \vp_j=0
$$
and the $c_j$'s we get the monodromy matrices in table~5.1.

\goodbreak
\midinsert
\vskip5mm
\vbox{
$$\vbox{\offinterlineskip
\hrule height 1.1pt
\halign{&\vrule width 1.1pt#&\strut\quad#\hfil\quad&
\vrule#&\strut\quad#\hfil\quad&\vrule width 1.1pt#\cr
height7pt&\omit&&\omit&\cr
&
$A~=~\left( \mutrix{\ph0 & \ph1 & \ph0 & \ph0 &\ph0 & \ph0 \cr
                   \ph0 &  \ph0 &  \ph1 & \ph0 &\ph0 & \ph0 \cr
                   \ph0 &  \ph0 & \ph0 & \ph1 & \ph0 & \ph0 \cr
                   \ph0 &  \ph0 & \ph0 & \ph0 & \ph1 & \ph0 \cr
                    \ph0 & \ph0 &  \ph0 & \ph0 &\ph0 & \ph1 \cr
                    -1 & -1 & -1 & -1 & -1 & -1  \cr}
                     \right)$

&&$T~~=~~
    \left(\mutrix{  \ph2 &  -1 &  \ph0 &  \ph0 &\ph0 &\ph0 \cr
                    \ph1 &  \ph0 &  \ph0 &  \ph0 &\ph0 &\ph0 \cr
                     -2&  \ph2 & \ph1 & \ph0 &\ph0 &\ph0 \cr
                     -1 &\ph1 & \ph0  & \ph1 &\ph0 &\ph0    \cr
                     \ph4 & -4& \ph0  & \ph0 & \ph1 & \ph0 \cr
                     -1 &  \ph1 &  \ph0 & \ph0 &\ph0 &\ph1 \cr} \right)$
&\cr
height7pt&\omit&&\omit&\cr} \hrule height
1.1pt}$$ \vskip5pt
\noindent
{\bf Table 5.1}: Monodromy matrices $(A,~T)$
associated to $\j=0,~(2^22^2)^{-1/7}$ for ${\cal W}_1$.}
\vskip2pt
\endinsert

As noted in Example~1 of section~2, by~(A.3) we have
$k=2$. The monodromy calculation then
gives
\eqn\eXXX{
U=T_{\infty}^2 - I\,,\qquad U^3=14\  E~.
}
Thus, by the argument in section~5.1 $y_{\j\j\j}^{\scriptstyle {\rm lcs}}=14$.
Now comparing with
the intersection calculation in section~2, we see that the couplings
indeed agree.

\noindent
{\bf Example 2.}
Let $p_2=0$, where $p_2$ is a transverse polynomial, define
a hypersurface ${\cal M}_2\in\cp{(1,1,3,4,6)}{4}[15]^{7,106}_{-198}$.
{}From \ePERF\ the period
corresponding to the fundamental deformation $\prod_{i=1}^5y_i$
on the mirror manifold ${\cal W}_2$ is
\eqn\eEXTWO{\vp_0~=~\sum_{m=0}^{\infty}{\G(15  m+1)\over \G^2(m+1)
 \G(3m+1)\G(4m+1)\G(6m+1) (15\j)^{ 15\  m}}.}
{}From~(B.16) we have  $c_j=(1,1,-1,0,-1,0,1,-1,2,-1,1,0,-1,0,-1)$
and there will
be $15-3=12$ linearly independent
periods.\foot{From~(B.8) we see that $k_3$ will result in three constraints
on the $\vp_j$ of the form $\sum_{j=0}^4 \vp_{l+3j}=0$ where $l=0,1,2$. There
will be no new restrictions from the other $k_i$.}
 Thus, in this case we
do not obtain all of the periods. This is a reflection of the fact that only
five of the total seven complex deformations are given as polynomial
deformations. This may provide important information in constructing ${\cal
W}_2$.  From the $c_j$'s above and the relation $\sum_{j=0}^4\vp_{3j}=0$ one
easily finds the relevant monodromy matrices, see table~5.2.
We
find
that $k=12$ and so
\eqn\eXXX{
U=T_{\infty}^{12} - I\,\qquad U^3=360\  E~.
}
Just as for the first example we conclude that
$y_{\j\j\j}^{\scriptstyle {\rm lcs}}=360$.
Comparing with the calculation done on ${\cal M}_2$ the couplings agree.

\goodbreak
\midinsert
\vbox{
$$\vbox{\offinterlineskip
\hrule height 1.1pt
\halign{&\vrule width 1.1pt#&\strut\quad#\hfil\quad&
\vrule  width 1.1pt#\cr
height7pt&\omit&\cr
&
$A~=~\left( \mutrix{
                     \ph0 & \ph1 & \ph0 & \ph0 &\ph0 & \ph0 &
                     \ph0 & \ph0 & \ph0 & \ph0 &\ph0 & \ph0  \cr
                     \ph0 &  \ph0 & \ph1 & \ph0 &\ph0 & \ph0 &
                     \ph0 & \ph0 & \ph0 & \ph0 &\ph0 & \ph0\cr
                      \ph0 &  \ph0 & \ph0 & \ph1 &\ph0 & \ph0 &
                     \ph0 & \ph0 & \ph0 & \ph0 &\ph0 & \ph0\cr
                      \ph0 &  \ph0 & \ph0 & \ph0 &\ph1 & \ph0 &
                     \ph0 & \ph0 & \ph0 & \ph0 &\ph0 & \ph0\cr
                      \ph0 &  \ph0 & \ph0 & \ph0 &\ph0 & \ph1 &
                     \ph0 & \ph0 & \ph0 & \ph0 &\ph0 & \ph0\cr
                      \ph0 &  \ph0 & \ph0 & \ph0 &\ph0 & \ph0 &
                     \ph1 & \ph0 & \ph0 & \ph0 &\ph0 & \ph0\cr
                      \ph0 &  \ph0 & \ph0 & \ph0 &\ph0 & \ph0 &
                     \ph0 & \ph1 & \ph0 & \ph0 &\ph0 & \ph0\cr
                      \ph0 &  \ph0 & \ph0 & \ph0 &\ph0 & \ph0 &
                     \ph0 & \ph0 & \ph1 & \ph0 &\ph0 & \ph0\cr
                      \ph0 &  \ph0 & \ph0 & \ph0 &\ph0 & \ph0 &
                     \ph0 & \ph0 & \ph0 & \ph1 &\ph0 & \ph0\cr
                      \ph0 &  \ph0 & \ph0 & \ph0 &\ph0 & \ph0 &
                     \ph0 & \ph0 & \ph0 & \ph0 &\ph1 & \ph0\cr
                      \ph0 &  \ph0 & \ph0 & \ph0 &\ph0 & \ph0 &
                     \ph0 & \ph0 & \ph0 & \ph0 &\ph0 & \ph1\cr
                       -1 & \ph0 & \ph0 & -1 &\ph0 &  \ph0 &
                       -1 & \ph0 & \ph0 & -1 &\ph0 &  \ph0 \cr}
                     \right)$&\cr
height8pt&\omit&\cr
\noalign{\hrule}
height8pt&\omit&\cr
&$T~~=~~
    \left(\mutrix{  \ph2 &  -1 &  \ph0 &  \ph0 &\ph0 &\ph0 &
                     \ph0 & \ph0 & \ph0 & \ph0 &\ph0 & \ph0  \cr
                     \ph1 &  \ph0 &  \ph0 & \ph0 &\ph0 &\ph0 &
                     \ph0 & \ph0 & \ph0 & \ph0 &\ph0 & \ph0  \cr
                      -1&  \ph1 & \ph1 & \ph0 &\ph0 &\ph0 &
                     \ph0 & \ph0 & \ph0 & \ph0 &\ph0 & \ph0  \cr
                      \ph0 &\ph0 & \ph0  & \ph1 &\ph0 &\ph0 &
                     \ph0 & \ph0 & \ph0 & \ph0 &\ph0 & \ph0  \cr
                      -1 & \ph1 & \ph0  & \ph0 &\ph1 &\ph0  &
                     \ph0 & \ph0 & \ph0 & \ph0 &\ph0 & \ph0  \cr
                      \ph0 &\ph0 & \ph0  & \ph0 &\ph0 &\ph1 &
                     \ph0 & \ph0 & \ph0 & \ph0 &\ph0 & \ph0  \cr
                      \ph1 &-1 & \ph0  & \ph0 &\ph0 &\ph0   &
                     \ph1 & \ph0 & \ph0 & \ph0 &\ph0 & \ph0  \cr
                      -1 &\ph1 & \ph0  & \ph0 &\ph0 &\ph0   &
                     \ph0 & \ph1 & \ph0 & \ph0 &\ph0 & \ph0  \cr
                      \ph2 &-2 & \ph0  & \ph0 &\ph0 &\ph0   &
                     \ph0 & \ph0 & \ph1 & \ph0 &\ph0 & \ph0  \cr
                      -1 &\ph1 & \ph0  & \ph0 &\ph0 &\ph0   &
                     \ph0 & \ph0 & \ph0 & \ph1 &\ph0 & \ph0  \cr
                      \ph1 &-1 & \ph0  & \ph0 &\ph0 &\ph0   &
                     \ph0 & \ph0 & \ph0 & \ph0 &\ph1 & \ph0  \cr
                      \ph0 &  \ph0 &  \ph0 & \ph0 &\ph0 &\ph0 &
                     \ph0 & \ph0 & \ph0 & \ph0 &\ph0 & \ph1 \cr} \right)$
&\cr
height7pt&\omit&\cr} \hrule height
1.1pt}$$ \vskip5pt
\noindent
{\bf Table 5.2}: Monodromy matrices $(A,~T)$ associated to $\j=0,
\quad(3^34^46^6)^{-1/15}$ for ${\cal W}_2$.}
\vskip2pt
\endinsert

In comparing the topological couplings computed on the 7,555 models
listed in \rAR\ with $y_{\j\j\j}^{\scriptstyle {\rm lcs}}$
obtained through monodromy
considerations described above we find complete agreement; though for a few
of the models $y_{\j\j\j}^{\scriptstyle {\rm lcs}}$ remains to be calculated.
This gives
a very large set of verifications of mirror symmetry,
 although only to lowest order.  The toric topological couplings have been
computed in principle for toric hypersurfaces up to a constant \rBatquant; our
work fixes the constant, verifying more of the conjectured mirror symmetry
in this situation.  In fact, we have shown that our methods apply to
a larger class of models than we have discussed above,
by checking agreement between the couplings on
$\cM$ and ${\cal W}$ for some non-transverse spaces considered in \rCOK.

In the above we have used the monomial-divisor mirror map~\rmdmm\
in finding $k$. This in turn was used to  show that the relevant monodromy
 is $T_\infty^k$ around the large complex structure limit point.
However, the actual numerical computation does not rely on these facts.
In a case by case study one finds
that $T_\infty$ has eigenvalues which are $k$th roots of unity. From this
and the general discussion in~\rCdFKM\ regarding the Yukawa couplings
in the large complex structure limit we deduce that the monodromy around the
large complex structure limit point indeed is $T_\infty^k$. The
definition of the flat coordinate $t$  then naturally follows. Thus all
the ingredients necessary for computing $y_{\j\j\j}^{\scriptstyle {\rm lcs}}$
can be found without relying on the toric picture in general and the
monomial-divisor mirror map in particular. It is very reassuring though, that
the two approaches agree and hence our explicit calculation lends further
support to some of the conjectures made in~\rmdmm.

\newsec{Discussions}\noindent
In this paper we have shown the agreement between the
$(1,1)$ form Yukawa coupling on ${\cal M}$ and the $(2,1)$ form
coupling computed on its mirror partner ${\cal W}$ in the large radius and
complex structure limit respectively, restricting to a particular
one-dimensional subspace of the moduli space. On the one hand this confirms
mirror symmetry, in a particular limit, for a large class of manifolds. On the
other hand the formulae for the respective coupling are very similar. This may
indicate that we ought to be able to formulate the computation only in terms of
the K\"ahler modulus. Unfortunately, although special geometry applies as well
for the K\"ahler class, very little is known about how we would carry out such
a calculation in practice.

The computation also shows that it is possible to find  the
Yukawa couplings and part of the modular group without knowing the integrable
basis, or at least with very little knowledge of it. This basis is in general
hard to construct and hence it is gratifying that detailed information about
the moduli space still can be obtained.

Finally (and this was one of our original motivations), we may view the
calculation as really being a calculation for a pair of mirror
families,
namely ${\cal M}\in\cp{(k_1,k_2,k_3,k_4,k_5)}{4}[d]$ which has just
one K\"ahler modulus, and the one parameter fundamental deformation of
${\cal W}$ which
we have been considering.  Much of the toric calculation was used merely
to justify the intuitive assertion that since we have to multiply the
naive K\"ahler class by $k$ to get an integral class, we must raise the
naive monodromy to the power $k$ to correctly scale the flat coordinate
for the fundamental deformation.  This complication arises because we are
dealing with Calabi-Yau {\it orbifolds} rather than manifolds.
The other main conclusion of the toric
calculation is the local description of the moduli space near $\j=\infty$.
While we used the methods of \rmdmm\ for this, we could have reached the
desired conclusion without this.  The advantage of organizing our
calculations in
this way is to also bring out the point that
we have explicitly checked some consequences
of the conjectures made in \rmdmm, as well as to make more precise the
conclusions of \rBatquant.

\noindent
{\bf Acknowledgements}:
It is a pleasure to thank P.~Candelas, X.~de la Ossa, T.~H\"ubsch,
F.~Liu, C.~Myers and
in particular D.~Morrison for useful discussions.
P.~B. was supported by the American-Scandinavian Foundation, the Fulbright
Program, NSF grants PHY 8904035 and PHY 9009850,
DOE grant DE-FG02-90ER40542 and the Robert~A.~Welch Foundation.
S.~K. was supported by NSF grant DMS-9311386.
P.~B. would also like to thank the ITP, Santa Barbara and the Theory Division,
CERN for their hospitality where part of this work was carried out.

\vfill
\eject

\appendix{A}{The Hyperplane Class}
Let us now turn to discuss  the hyperplane class $H$ of a weighted hypersurface
$\cM$ in $\wp$.
The map $\r:\IP^4\to\wp$
induces an inclusion $\r^*:H^2({\cal M},\ZZ)\to H^2(X,\ZZ)$ with $X \subset
\IP^4$.  Let
${\tilde H}\in H^2(X,\ZZ)$ be the hyperplane class induced from the one on
$\IP^4$.
Let $H\in H^2({\cal M},\ZZ)$ be the unique generator with the
property that $\r ^*(H)$ is positive.  Suppose that $\r^*(H)=k\tilde{H}$.
We summarize this situation by saying that $k$ is the weight of the
hyperplane class $H$ of ${\cal M}$.  $k$ is
all that
is needed to calculate the intersection numbers that we need.  In fact:
\eqn\egen{
H^3=\frac{k^3d}{\prod_{i=1}^5k_i}~.}
(Compare with \rHKTY).

To see this,
note that since $\r$ is finite of degree $\prod_{i=1}^5k_i$, we have
\eqn\egenp{
H^3=\frac1{\prod_{i=1}^5k_i}(\r^*(H))^3=\frac{k^3d}{\prod_{i=1}^5k_i}~,}
the last equality holding since the generator of $H^2(X,\ZZ)$ has degree $d$
(i.e.\ $X$ has degree $d$).

It remains to find the weight $k$ of $H$.
Equivalently, this is the smallest
positive integer with the property that the restriction to ${\cal M}$ of forms
 of
degree $k$ are identified with the sections of a line bundle.  It suffices
to show this when the forms are restricted to each of a collection of open
subsets of ${\cal M}$, which taken together cover ${\cal M}$.
Equivalently, there must exist on sufficiently small open sets a nowhere
vanishing holomorphic form of degree $k$
(which is equivalent to giving a trivialization
of the bundle).
For generalities on the pathologies
that can occur, see \rdoltor.

We first illustrate with a simple example,
 $\WCP{4}{(1,1,2,2,2)}[8]^{2,86}_{-168}$ \rCdFKM.
By a {\it rational monomial}, we mean an expression $\prod_ix_i^{r_i}$, where
$r_i$ is an integer (allowed to be negative).
We claim that $k=2$.  First note that $k=1$ does not suffice.
To see this, first note that the only rational
monomials of degree $1$ must contain terms involving $x_0$ and/or $x_1$ in the
numerator or denominator. These monomials cannot be both holomorphic and
nowhere vanishing along the locus $x_0=x_1=0$, which is a plane quartic
curve.  On the other hand, for $k=2$, we consider the open cover
$U_i=\{(x_1,\ldots,x_5)|x_i\neq 0\}$, and exhibit the respective degree $2$
 forms
$x_1^2,x_2^2,x_3,x_4,x_5$ which are holomorphic and nowhere vanishing in
the respective open sets $U_1,\ldots,U_5$.

Another general feature exhibited by this example is that the problem
occurs along the singular locus (the plane quartic), which has quotient
singularities of order $2$.  As we shall see presently, $k=2$ can be derived
from this fact.

We now turn to the general case.  First of all, if none of the $x_i$ are
zero (so that we are situated in the smooth locus), then any degree
suffices locally: since gcd$(k_1,\ldots,k_5)=1$, a rational monomial of
degree $1$ can be constructed, and any such monomial may be used to
trivialize a
bundle.  So the only problems can occur at points at which at least one
of the coordinates is $0$.

Without loss of generality, assume that we are at a point where
$x_1=x_2=\ldots =x_s=0$
while all other coordinates are non-zero.
We must find a rational expression of degree $k$
with neither
zeros nor poles at this point.  If such an expression exists, we can
substitute $x_1=x_2=\ldots =x_s=0$ into it to get another.  In this way,
we may assume that such an expression only involves $x_{s+1},\ldots,x_5$.
For this to be possible, the gcd of $k_{s+1},\ldots,k_5$ must divide $k$.
Conversely, if the gcd of $k_{s+1},\ldots,k_5$ divides $k$, then a degree
$k$ rational monomial can be constructed from the last $5-s$ coordinates, and
such a monomial is holomorphic and non-vanishing in a neighborhood of the
point in question.

Now we investigate which combinations of coordinates can be zero for some
point of $\Mhat$.  This is easy.  Setting three or fewer coordinates to 0
gives a positive dimensional subspace of $\wp$, hence the hypersurface
${\Mhat}$ must
have non-empty intersection with it.  On the other hand, if four coordinates
are set to zero (say the first $4$), then this determines the unique point
$(0,0,0,0,1)$ of $\WP^4$.  This point is in ${\Mhat}$ (we assume ${\Mhat}$ is
general)
if and only if every degree $k$ monomial involves at least one of the first
$4$ coordinates, or equivalently, there is no monomial involving only $x_5$.
This is equivalent to $d$ not being a multiple of $k_5$.
Thus, we see that $k$ is given by
\eqn\ekkk{
k~=~\hbox{lcm}(\,\{\gcd(k_i,k_j)\,| \quad i\neq j\ \forall i,j\}~\cup~
                \{k_i\,| \quad k_i\ \hbox{does not divide }d\}\,)~.
}

Alternatively, one can use the notion of Cartier divisors rather than line
bundles.  The point is that the zero locus of a form of degree $k$ cannot
be a Cartier divisor near a point of the singular locus unless its degree
is a multiple of the index of the singular point.
By the index, we mean the order of the isotropy group of a point of $X$ lying
over the point in question.  This is the approach taken in a similar
calculation appearing in \rHKTY.

It is easy to see that the calculation works out exactly as above.  The
index of a point where $x_1=\ldots =x_s=0$ is just the gcd of
$k_{s+1},\ldots,k_5$ as before, as one easily computes.  This includes
smooth points, which have index $1$.

\appendix{B}{Monodromy}%Section 4
\noindent
In what follows we will use the technique developed in \rCdGP\ to obtain the
monodromy matrices around the singular points in the modular space of
$\j$-deformations. For further details we refer the reader to \rCdGP.

By studying the condition under which $\vp_0$ converges one finds that there
is a conifold singularity\foot{Strictly speaking we would have to
compute the matrix of second derivatives of $\hat p_\j$ to show that it is
not singular  where $\hat p_\j=\rd \hat p_\j=0$. For now
we will assume that this is a singularity of conifold type and return to this
question as we discuss the  monodromy around the singularity.}
 given by
\eqn\eXXX{
          1 - \prod_{i=1}^5k_i^{k_i}\j^d~=~0~.
}
 In order to find the monodromy around this singularity we need to analytically
continue \ePER\ to small $\j$. Let us assume that $k_1$ is the smallest
weight.
By using the multiplication formula,
\eqn\eMULG{
   \G(k_1n+1)~=~n\,k_1^{k_1n+1/2} (2\p)^{(1-k_1)/2}
               \prod_{r=0}^{k_1-1} \G(n+{r\over k_1})
}
we can rewrite \ePER\ as
\eqn\ePERFF{\hbox{}\mkern-20mu
\eqalign{\vp _0~&=~(2\p)^{(k_1-1)/2} k_1^{-1/2}\cr
  &\times\sum_{n=0}^{\infty} {\G(d\  n +1)\over \G(n+1)
  \prod_{r=1}^{k_1-1} \G(n+{r\over k_1}) \prod_{i=2}^5 \G(k_i\  n+1)
  (k_1^{k_1}(d\j)^d)^n}~.}
}
To analytically continue $\vp_0$ to small $\j$ we use Barnes integral to write
\ePERFF\ as
\eqn\eBAR{\vp _0~=~{(2\p)^{(k_1-1)/2}\over 2\p i k_1^{1/2}}
\int_{\g} \rd s{\G(-s) \G(d\,s+1) e^{i\p s} (k_1^{k_1/d}d\j)^{-ds} \over
\prod_{r=1}^{k_1-1}\G(s+{r\over k_1}) \prod_{i=2}^5\G(k_is+1)}~.}

By closing $\g$, given by $s=-\e+i y$, where  $y \in (-\infty,+\infty)$ and
$0<\e<1/d$, to the right and demanding $|\j|>(\prod_{i=1}^5k_i^{k_i/d})^{-1}$
 we recover \ePERFF\ due
to the poles in $\G(-s)$.
However, there are also poles for $d\,  s+1=-m$ with $m=0,1,\ldots$ and by
requiring that $|\j|<(\prod_{i=1}^5k_i^{k_i/d})^{-1}$ we can encircle the
latter poles by closing the
contour to the left,
\eqn\ePERFA{
\eqalign{\vp _0~&=~-{(2\p)^{(k_1-1)/2}\over k_1^{1/2}d}\cr
&\times ~\sum_{m=1}^{\infty} {\G({m\over d})\a^{{d-1\over2}m}
(k_1^{k_1/d}d\j)^m\over \G(m)\prod_{r=1}^{k_1-1}\G({r\over k_1} -{m\over d})
\prod_{i=2}^5 \G(1-{k_im\over d})}~.}
}

So far we have only considered one period. But we know on general grounds that
$\vp_0$ satisfies a generalized hypergeometric equation of order $q$ to which
there are $q$ linearly independent solutions. To construct the other $q-1$
solutions we use that the generator of the phase symmetry in the modular group
is given by
\eqn\ePHASE{{\cal A}~:~\j \to \a \j~~,~~\a^d=1~.}
This follows from $\W(\a\j)=\W(\j)$ \rCdGP, \ie\ the theory is invariant
under the action of ${\cal A}$.
So $\vp_0(\a\j)$ must also be a solution to the Picard-Fuchs equation. Hence,
we
obtain the other periods by acting with ${\cal A}$,
\eqn\ePERFJ{
\eqalign{\vp _j(\j)&
\define \vp _0(\a^j\j) = -{(2\p)^{(k_1-1)/2}\over k_1^{1/2}d} \cr
&\times \sum_{m=1}^{\infty} {\G({m\over d})\ \a^{({d-1\over2}+j)m}
 (k_1^{k_1/d}d\j)^m\over
\G(m)\prod_{r=1}^{k_1-1}\G({r\over k_1} -{m\over d}))
\prod_{i=2}^5 \G(1-{k_im\over d})}~.  \cr}
}
Note that the $\vp_j$ are not all linearly independent. This is seen
by observing
that the denominator has poles for
\eqn\eREL{
     {m\  k_i\over d}~=~1,2,\ldots~~{\rm and}~~{m\over d}-
                            {r\over k_1}~=~0,1,\ldots~.}
In particular we have the relation
\eqn\eRELALL{\sum_{j=0}^{d-1} \vp _j~=~0~,}
as one relation but depending on the weights $k_i$ there will in general
be more. By using relations of the type \eRELALL\ we can write down the
matrix, $A$, associated to the action of ${\cal A}$. Then, defining the period
vector as
\eqn\ePERV{{\bf \vp }~=~\left( \mutrix{\vp _0 \cr
                                     \vp _1\cr
                                     \vdots \cr
                                     \vp _{q-1}\cr}
                     \right)~,}
the action of  ${\cal A}$ is simply given by (see eq.~\ePHASE\ )
\eqn\eACTA{{\cal A}~: \vp _j \to \vp _{j+1}~,~~~ j=0,\ldots,q-1~.}
But $\vp_q$ is not one of the components of ${\bf \vp }$. However,
\eqn\eWP{\vp _q~=~\sum_{j=0}^{q-1} a_j \vp _j}
where the $a_j$ are determined by relations like \eRELALL\ .

The next step is to compute the action on the $\vp_j$ when going around the
conifold $\j=(\prod_{i=1}^5k_i^{k_i/d})^{-1}$. (Due to the phase symmetry
${\cal A}$ the $d$ different
 singularities are identified and it is enough to study the above mentioned
one.) Under transport around the singularity, $\vp_j$ transforms
according to \rCdGP\
\eqn\eTRAN{\vp_j(\j) \to \vp _j(\j) + c_j z^0(\j),}
where $z^0=\vp _0-\vp _1$  vanishes at the singularity like
$(\j-(\prod_{i=1}^5k_i^{k_i/d})^{-1})$. Equivalently,
\eqn\ePERC{\vp_j~=~{c_j\over 2\p i}z^0(\j)
 \log(\j-(\prod_{i=1}^5k_i^{k_i/d})^{-1}) + \ldots~,}
where the ellipses indicate terms analytic in a neighborhood of the conifold.
Thus, the $c_j$
can be obtained by studying $\vp_j$ for large $m$ in \ePERFJ\ along the lines
of ref.~\rCdGP\ . We get
\eqn\eCJ{\hbox{}\mkern25mu
\eqalign{c_j&\define \sum_{s=0}^{d-1}c_j(s) \cr
\phantom{c_j}&={1\over d}e^{\p i (1-k_1)}\sum_{s=0}^{d-1}\a^{js}
 \prod_{r=1}^{k_1-1} (\a^s e^{-2\p i r/k_1} - 1)
\prod_{i=2}^5 (\a^{s k_i} - 1)~.}
}
Thus, \ePERC\ and \eCJ\ show that $\j=(\prod_{i=1}^5k_i^{k_i/d})^{-1}$ indeed
is a conifold singularity.
Eq.~\eCJ\ can be simplified slightly by expanding
$\prod_{r=1}^{k_1-1} (\a^s e^{-2\p i r/k_1} - 1)$ giving us the final
expression
\eqn\eCJF{
\eqalign{c_j&={1\over d}\sum_{s=0}^{d-1}\a^{js} \sum_{l=0}^{k_1-1}\a^{sl}
\prod_{i=2}^5 (\a^{s k_i} - 1) \cr
&=\sum_{l=0}^{k_1-1}\sum_{r=0}^4\sum_{i\in(2,\ldots,5)_r}\left[{k_i
+l+j\over d}\right]^r~.}
}
The last form is especially useful since it shows the integral structure
and hence is preferable for actual computation, see examples in section~5.

Note that for $k_i=1~,~i=1,\ldots,5$ we recover the expression obtained in
\rCdGP\ . It is also worth pointing out that by further studying \eCJF\ one
can show that $c_0=c_1=1$.
Thus, for any given model we are now able write down both the monodromy
matrix, $T$, associated to transport around
$\j=(\prod_{i=1}^5k_i^{k_i/d})^{-1}$ using \eTRAN\ and \eCJF\
as well as the phase symmetry matrix, $A$,  see table B.1. Finally, the
monodromy  around $\j=\infty$  is given by \rCdGP\
\eqn\eTINF{T_{\infty}~=~(A\, T)^{-1}~.}

\goodbreak
\midinsert
\vbox{
$$\vbox{\offinterlineskip
\hrule height 1.1pt
\halign{&\vrule width 1.1pt#&\strut~#\hfil~&
\vrule#&\strut~#\hfil~&\vrule width 1.1pt#\cr
height7pt&\omit&&\omit&\cr
&
$A=\left( \mutrix{\ph0 & \ph1 & \ph0 &\ \ldots & \ph0 \cr
                   \ph0 &  \ph0 &  \ph1&\ \ldots & \ph0 \cr
                 \ph\vdots &\ph\vdots  &\ph\vdots &\ \ddots&\ph\vdots     \cr
                    \ph0 & \ph0 &  \ph0 &\ \ldots & \ph1 \cr
                    \ph a_0 & \ph a_1 & \ph a_2 &\ \ldots & \ph a_{q-1} \cr}
                     \right)$

&&$T=
    \left(\mutrix{  \ph2 &  -1 &  \ph0 &  \ph0 &\ \ldots &\ph0 \cr
                    \ph1 &  \ph0 &  \ph0 &  \ph0 &\ \ldots &\ph0 \cr
                    \ph c_2 &  -c_2 & \ph1 & \ph0 &\ \ldots &\ph0 \cr
            \ph\vdots &\ph\vdots &\ph\vdots &\ph\vdots &\ \ddots &\ph\vdots \cr
          \ph c_{q-1} &  -c_{q-1} &  \ph0 & \ph0 &\ \ldots &\ph1 \cr} \right)$
&\cr
height7pt&\omit&&\omit&\cr} \hrule height
1.1pt}$$
\noindent
{\bf Table B.1}: Monodromy matrices $(A,~T)$ associated to
 $\j=0,\,(\prod_{i=1}^5k_i^{k_i/d})^{-1}$.
The number of linearly independent periods is $q$.}
\endinsert

\vfill
\eject

\listrefs

\bye